\documentclass[%
 aip,
 amsmath,amssymb,
 reprint,%
]{revtex4-1}

\usepackage{graphicx}
\usepackage{dcolumn}
\usepackage{bm}

\usepackage[utf8]{inputenc}
\usepackage[T1]{fontenc}
\usepackage{mathptmx}
\usepackage{etoolbox}
\usepackage{xcolor}
\usepackage{hyperref} 
\usepackage{multirow}
\usepackage{amsmath}
\usepackage{upgreek}
\usepackage[flushleft]{threeparttable}

\makeatletter
\def\@email#1#2{%
 \endgroup
 \patchcmd{\titleblock@produce}
  {\frontmatter@RRAPformat}
  {\frontmatter@RRAPformat{\produce@RRAP{*#1\href{mailto:#2}{#2}}}\frontmatter@RRAPformat}
  {}{}
}%
\makeatother
\begin{document}

\preprint{AIP/123-QED}
\title{The influence of non-Newtonian behaviors of blood on the hemodynamics past a bileaflet mechanical heart valve}
\author{A. Chauhan}
\author{C. Sasmal*}
 \email{csasmal@iitrpr.ac.in}
\affiliation{Department of Chemical Engineering, Indian Institute of Technology Ropar, Rupnagar, Punjab, India-140001}

\date{\today}

\begin{abstract}

This study employs extensive three-dimensional direct numerical simulations (DNS) to investigate the hemodynamics around a bileaflet mechanical heart valve. In particular, this study focuses on assessing whether non-Newtonian rheological behaviors of blood, such as shear-thinning and yield stress behaviors, exert an influence on hemodynamics compared to the simplistic Newtonian behavior under both steady inflow and physiologically realistic pulsatile flow conditions. Under steady inflow conditions, the study reveals that blood rheology impacts velocity and pressure field variations, as well as the values of clinically important surface and time-averaged parameters like wall shear stress (WSS) and pressure recovery. Notably, this influence is most pronounced at low Reynolds numbers, gradually diminishing as the Reynolds number increases. For instance, surface-averaged WSS values obtained with the non-Newtonian shear-thinning power-law model exceed those obtained with the Newtonian model. At $Re = 750$, this difference reaches around 67\%, reducing to less than 1\% at $Re = 5000$. Correspondingly, pressure recovery downstream of the valve leaflets is lower for the shear-thinning blood than the constant viscosity one, with the difference decreasing as the Reynolds number increases. On the other hand, in pulsatile flow conditions, jets formed between the leaflets and the valve housing wall are shorter than steady inflow conditions. Additionally, surface-averaged wall shear stress and blood damage (BD) parameter values are higher (with differences more than 13\% and 47\%, respectively) during the peak stage of the cardiac cycle, especially for blood exhibiting non-Newtonian yield stress characteristics compared to the shear-thinning or constant viscosity characteristics. Therefore, blood non-Newtonian behaviors, including shear-thinning and yield stress behaviors, exert a considerable influence on the hemodynamics around a mechanical heart valve. All in all, the findings of this study demonstrate the importance of considering non-Newtonian blood behaviors when designing blood-contacting medical devices, such as mechanical heart valves, to enhance functionality and performance.

\end{abstract}

\maketitle

\section{\label{into}Introduction}
Human blood, an intricate bodily fluid coursing through the cardiovascular system, serves as a conduit for vital substances across the body. Comprising diverse elements like red blood cells (RBCs), white blood cells (WBCs), platelets, and plasma, human blood diverges significantly from the simplicity of Newtonian fluids, such as water, wherein stress and deformation rate adhere to a linear relationship as defined by Newton's law of viscosity. Instead, it manifests intricate non-linear relationships, marking it as a complex non-Newtonian suspension. Consequently, human blood exhibits a spectrum of non-Newtonian behaviors in its rheological characterization. Predominantly, human blood demonstrates shear-thinning behavior, where viscosity diminishes with increasing deformation rates, a phenomenon established decades ago~\citep{chien1970shear,baskurt2011red}. However, ongoing research has uncovered several more intricate rheological behaviors of human blood~\citep{beris2021recent}. Notably, human blood exhibits both viscous and plastic behaviors, typified by viscoplastic tendencies featuring a discernible yield stress~\citep{picart1998,morris1989evaluation}. Furthermore, it displays viscoelastic behaviors arising from the deformation of red blood cells and the formation of reversible aggregates out of them, known as rouleaux aggregates~\citep{campo2013viscoelasticity}. Surprisingly, recent investigations reveal that even blood plasma, primarily composed of water (90\%) and other components like dissolved proteins, glucose, electrolytes, and hormones, also exhibits viscoelastic behavior~\citep{varchanis2018viscoelastic,brust2013rheology}. Moreover, human blood demonstrates the more intricate thixotropic rheological behavior, characterized by time-dependent shear-thinning and elastic properties, attributed to the formation of rouleaux influenced by the Brownian motion of RBCs~\citep{huang1987thixotropic,thurston1979rheological,horner2019measurements}.

These intricate non-Newtonian rheological behaviors exhibited by human blood introduce an increased level of complexity into its flow dynamics, commonly referred to as hemodynamics, surpassing the simplistic assumptions of it being a Newtonian fluid~\citep{baskurt2007handbook,cokelet1980rheology,cokelet2011hemorheology}. Even within straight microvessels, the flow of human blood deviates from the anticipated characteristics of constant viscosity Newtonian fluids. Instead, it yields a blunt velocity profile with higher velocities near the vessel wall. This deviation arises from the localized changes in blood viscosity attributable to the deformation and formation of aggregates of red blood cells, diverging from the parabolic velocity profile observed in Newtonian fluids~\citep{yeom2014changes}. These nuanced hemodynamic phenomena hold substantial implications for calculating critical clinical parameters, including flow resistance and hemodynamic shear stress, often interchangeably known as wall shear stress (WSS). Accurate determination of these parameters plays a pivotal role in diagnosing cardiovascular diseases, particularly atherosclerosis, and informs the development of intervention and treatment strategies~\citep{davies2009hemodynamic}. Beyond disease identification, an understanding of blood rheology proves crucial in the design and fabrication of medical devices, especially those in direct contact with blood. This is essential for ensuring their precise functionality and operation. Importantly, the consideration of blood rheology becomes imperative in mitigating the risk of material thrombosis and subsequent catastrophic adverse thrombotic events, such as stroke or pulmonary embolism, associated with the formation of blood clots on these devices~\citep{hong2020evaluating}. 

The prosthetic mechanical heart valve is one such medical device that makes direct contact with blood. These medical devices are crucial in addressing valvular heart diseases and are in high demand globally due to the prevalence of these diseases affecting millions each year~\citep{zilla2008prosthetic,vongpatanasin1996prosthetic}. Conditions like congenital abnormalities, endocarditis, atherosclerosis, high blood pressure, aging, and specific diseases contribute to the destruction of native heart valves, necessitating replacement~\citep{DeSilva2013, Liu2015}. The failure of heart valves to open and close properly, a consequence of diseases, leads to inefficient blood pumping, resulting in heart failure, cardiac arrest, and death. Cardiovascular diseases, responsible for 32\% of global mortality, claim 85\% of their victims due to heart attacks~\citep{cdc}. To address valvular heart diseases, prosthetic heart valves, whether mechanical or bioprosthetic, aim to replicate native valve functions while ensuring smooth hemodynamics, long durability, thromboresistance, and implantability~\citep{Vongpatanasin1996}. However, available prosthetic heart valves have limitations; mechanical valves offer durability exceeding 25 years but exhibit poor thromboresistance, while bioprosthetic valves excel in thromboresistance but have a lifespan limited to around 15 years~\citep{Sun2009}. Consequently, mechanical valves are often preferred for patients under 60 years old due to their superior durability~\citep{Jaffer2016}.

The deficient thromboresistant capability of mechanical heart valves stems from hemolysis-induced damage to red blood cells and platelet activation, attributed to nonphysiological flow patterns in and around the valve~\citep{gott2003mechanical}. Presently available prosthetic mechanical heart valves, despite incorporating the latest designs, induce undesirable flow features such as jet flow, elevated shear stresses, flow separation, recirculation, shed vortices, and turbulence. These factors pose threats of damaging red blood cells, activating platelets, and fostering blood clot formation. Consequently, the meticulous design of mechanical heart valves is crucial, necessitating a comprehensive investigation into fluid-structure interaction (FSI) involving the valve leaflet, heart wall, and blood. Additionally, molecular-scale interactions between the valve surface and blood cells are pivotal considerations.

Therefore, a substantial body of literature, encompassing both experimental and numerical studies, delves into fluid mechanics pertaining to prosthetic mechanical heart valves. For instance, Chandran et al.~\citep{chandran1985experimental} experimentally examined physiological pulsatile flow patterns past a caged-ball valve (Starr-Edwards valve) in a model human aorta using laser Doppler anemometry (LDA). Their findings unveiled an asymmetric velocity profile downstream and a jet-like flow structure in the peripheral region of the valve. Subsequent investigations~\citep{chandran1985experimental2} explored a tilting disc valve (Bjork-Shiley valve), revealing a bi-helical secondary flow structure downstream complicated by the multiple curvatures of the aorta and disc orientation. Gross et al.~\citep{gross1988vortex} conducted detailed investigations on the flow dynamics of two bileafet valves, St. Jude Medical and Carbomedics, utilizing the particle image velocimetry (PIV) technique. They identified a von Karman-like vortex pattern, characterized by a pair of vortices shedding from the valve surface. Further exploration of vorticity dynamics in a bileafet valve by Dasi et al.~\citep{dasi2007vorticity} emphasized the significance of flow through the hinge region, revealing a vortex during forward flow, evolving into a disturbed three-dimensional structure during reverse flow with zones of high turbulent shear stresses capable of damaging red blood cells.

Comparative studies among different prosthetic valves, including in vitro assessments by Kvitting et al.~\citep{kvitting2010vitro}, elucidated distinct differences in velocity fields and turbulence kinetic energy. Beyond experimental studies, extensive numerical investigations provided insights into wall and turbulent shear stresses. For example, Thalassoudis et al.~\citep{thalassoudis1987numerical} numerically studied turbulent flow past a Starr-Edwards caged-ball valve, identifying maximum turbulent shear stresses near the sewing-ring tip and in the sinus separation region, consistent with experimental observations. Huang et al.~\citep{huang1994numerical} conducted a numerical study on a tilting disc valve, observing maximum shear stress at the valve disc, deemed insufficient to damage red blood cells. Cheng et al.~\citep{cheng2003two} simulated a bileafet valve, identifying relatively high velocity and shear stress fields in the clearance region between the leaflet and valve housing during valve closure, potentially contributing to blood clot formation.

Numerous studies employ diverse numerical tools to scrutinize the flow dynamics in various mechanical valves, with comprehensive summaries available in excellent review articles~\citep{sotiropoulos2009review,khalafvand2011cfd}. Furthermore, outstanding review articles comprehensively delineate the functionality, durability, hemodynamic ability, thromboresistant capability, development status, and future directions of different mechanical heart valves from fluid mechanical perspectives~\citep{yoganathan2004fluid,yoganathan2005flow,sotiropoulos2016fluid}. Intriguingly, all prior studies, whether experimental or numerical, have hitherto presumed blood as a simple Newtonian fluid. In contrast, as emphasized earlier, blood exhibits diverse non-Newtonian characteristics, encompassing shear-thinning, viscoplasticity, viscoelasticity, and more complex thixotropic rheological behavior. This raises the pivotal question: Do non-Newtonian blood behaviors significantly impact hemodynamics past a mechanical heart valve? This article endeavors to address this query through extensive three-dimensional direct numerical simulations (DNS) under both steady inflow and realistic physiological pulsatile flow conditions, considering various non-Newtonian behaviors of blood. Specifically, this study aims to meticulously investigate whether specific non-Newtonian behaviors of blood can, indeed, modify flow characteristics, such as turbulence intensity, vortex dynamics, blood damage, or forces acting on the leaflet, potentially influencing the performance of a mechanical heart valve. Consequently, this investigation seeks to furnish in-depth insights and analyses of these fluid mechanical aspects, contributing valuable information for designing next-generation mechanical heart valves with enhanced hemodynamic performance and thromboresistant capability.

\section{\label{ProbFor}Problem setup}

\begin{figure}
    \centering
    \includegraphics[trim=0cm 0cm 0cm 0cm,clip,width=8cm]{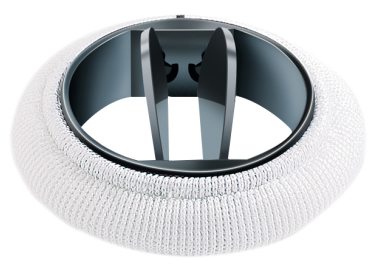}
    \caption{ St. Jude Medical heart valve (SJMHV).} 
    \label{SJMHV}
\end{figure}

\begin{figure*}
    \centering
    \includegraphics[trim=1.5cm 24.5cm 1cm 1cm,clip,width=16cm]{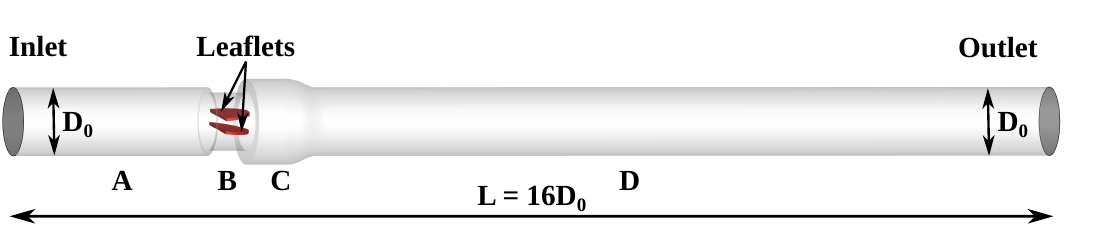}
    \caption{The whole computational domain used in the present numerical setup, which consists of four regions: \textbf{A}-ventricular side chamber, \textbf{B}-valve, \textbf{C}-sinus expansion and \textbf{D}-aortic side chamber. Note that the origin is set at the center of the inlet plane such that the $x$ axis is in the streamwise direction, the $y$ axis is in the spanwise direction, and the $z$ axis is parallel to the long axis of the leaflets.} 
    \label{Geometry}
\end{figure*}

\begin{figure}
    \centering
    \includegraphics[trim=0cm 0cm 0cm 0cm,clip,width=8cm]{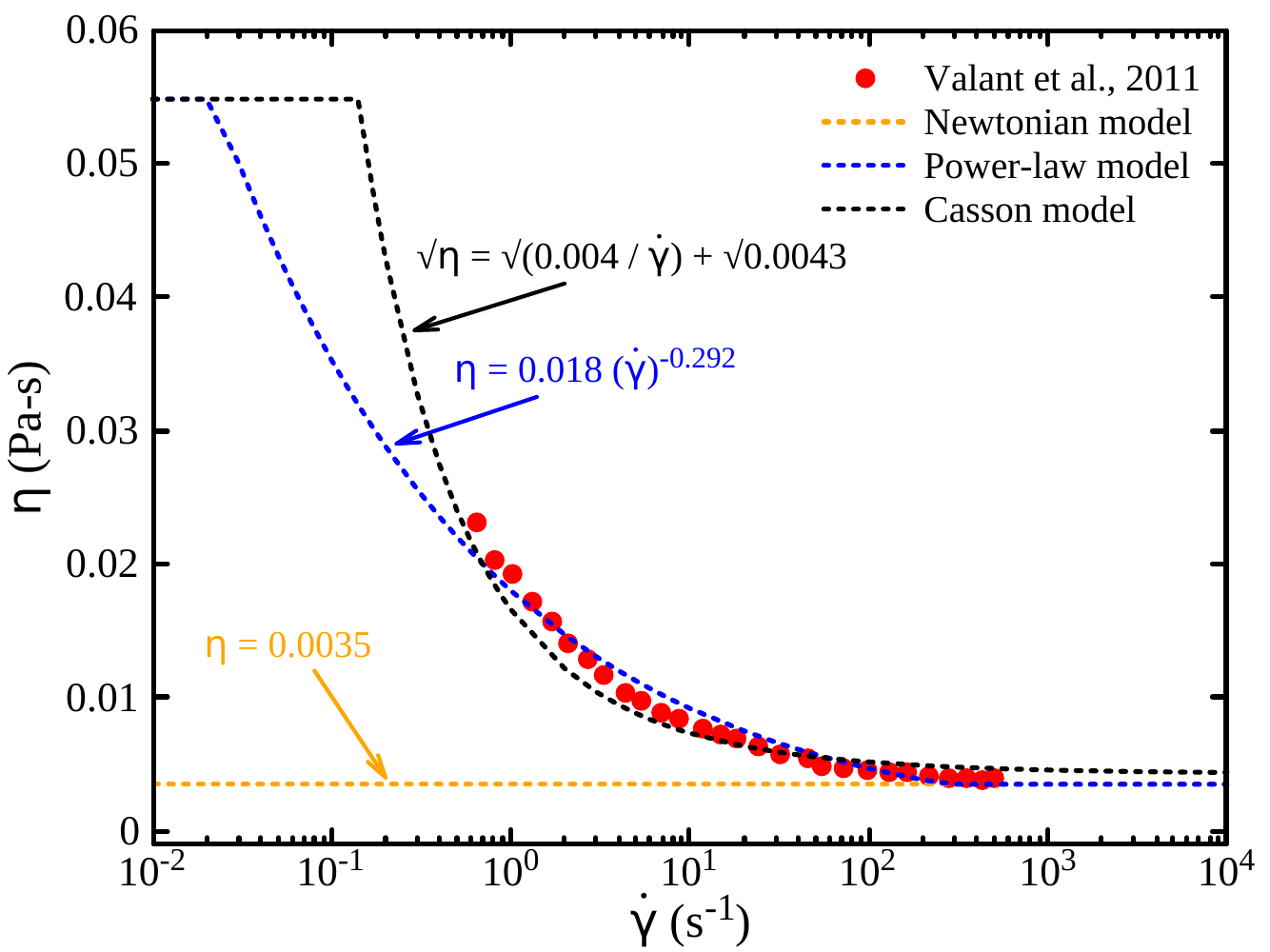}
    \caption{Shear viscosity fitting curves of blood with the experimental results obtained by Valant et al.~\citep{Valant2011} using different rheological models implemented in the present study.} 
    \label{ShearViscosity}
\end{figure}

\begin{figure}
    \centering
    \includegraphics[trim=0cm 0cm 0cm 0cm,clip,width=8cm]{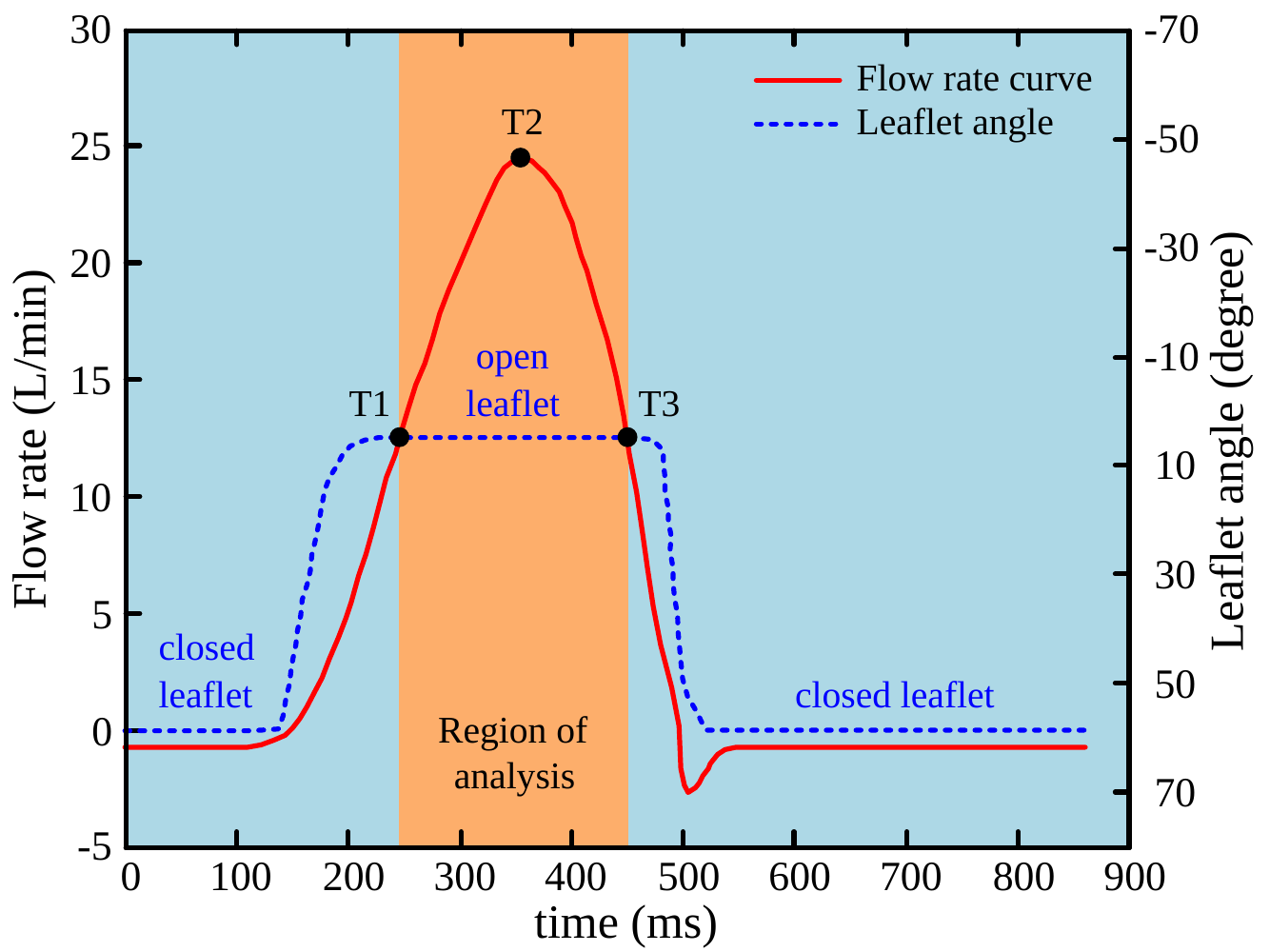}
    \caption{Variations of the flow rate (left axis) and leaflet angle (right axis) of the mechanical heart valve for one cardiac cycle of $860$ $ms$ corresponding to a heart rate of $70$ $beat/min$~\citep{Yun2014}. Here, the time instances T1, T2, and T3 denote the mid-acceleration, peak, and mid-deceleration phases during the systolic condition, respectively.} 
    \label{FlowCurve}
\end{figure}

\begin{figure*}
    \centering
    \includegraphics[trim=0.2cm 0cm 0cm 0.2cm,clip,width=16cm]{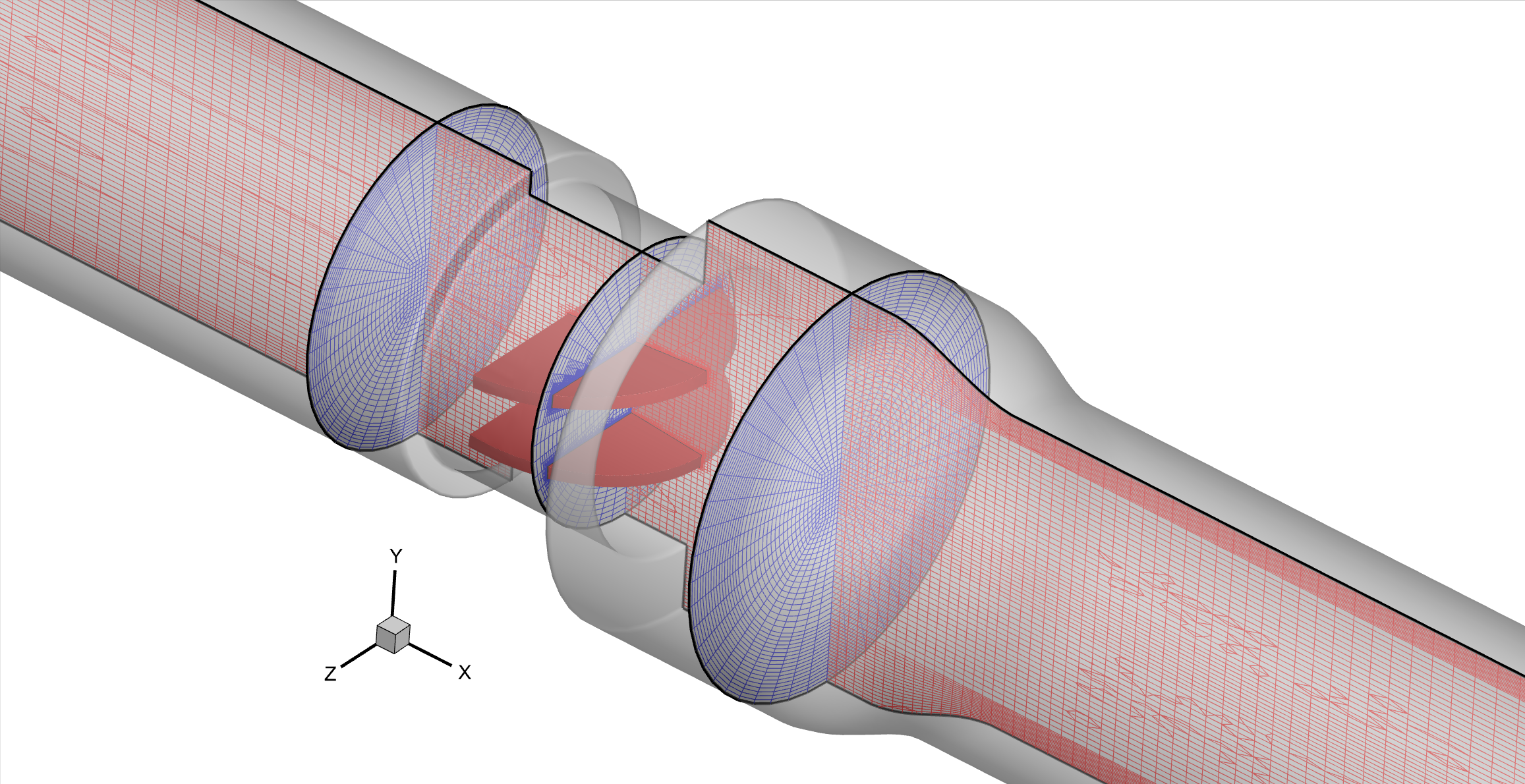}
    \caption{The schematic of grid 2 (G2) with a total of about 4.25 million cells used in the present study showing the refined mesh structure near the two leaflets and the solid walls to capture the steep gradients of velocity, pressure, stress, etc. } 
    \label{Mesh}
\end{figure*}

This study aims to investigate the influence of non-Newtonian behaviors of blood on the hemodynamics past mechanical heart valve, particularly a St. Jude Medical (SJM) mechanical heart valve with a diameter of 23 $mm$, as depicted in Fig.~\ref{SJMHV}. This valve is commonly employed for valve replacement in clinical practice~\citep{emery2005st,baudet1995long}. It has also been extensively utilized in previous research, both in simulations and experimental studies, to gain insights into the hemodynamics associated with mechanical heart valves~\citep{Ge2005, Yun2014}. The computational domain, illustrated in Fig.~\ref{Geometry}, consists of four distinct regions: the ventricular side chamber (A), valve (B), sinus expansion (C), and aortic side chamber (D). The valve's internal diameter is denoted as $D_B = 21.4$ $mm$, while the diameters of the aortic and ventricular sides are equal, each with a value of $25.4$ $mm$ ($D_A = D_D = D_0 = 25.4$ $mm$). The lengths of these chambers are specified as $L_A = 3D_0$ and $L_D = 11.35D_0$. The overall computational domain spans sixteen times the mean diameter, i.e., $L = 16 D_0$, and features a sudden axisymmetric expansion downstream of the valve (region C) with a diameter of $D_C = 31.75$ $mm$. We adopt a Cartesian coordinate system with the origin at the center of the inlet plane, the $x$ axis along the streamwise direction, the $y$ axis perpendicular, and the $z$ axis parallel to the long axis of the leaflets. The present structure of the computational domain aligns with previous work by Yun et al.~\citep{Yun2014}. Additionally, the blood flow is assumed to be incompressible, exhibiting non-Newtonian shear-thinning and yield stress behaviors. As determined by Valant et al.~\citep{Valant2011}, the non-Newtonian properties of blood are incorporated into the present simulations using the power-law and Casson fluid models, as depicted in Fig.~\ref{ShearViscosity}. Furthermore, Fig.~\ref{FlowCurve} illustrates the implemented volumetric flow rate and the dependency of leaflet angle throughout one cardiac cycle lasting 860 $ms$, corresponding to a heart rate of 70 beats/min, as reported by Yun et al.~\citep{Yun2014}. The peak and average flow rates for one cardiac cycle are approximately $25$ L/min and $4.5$ L/min, respectively. Our analysis focuses on a specific region where both leaflets are held fixed at a leaflet angle of $5^{\circ}$ on the $xz$-plane (neglecting the hinge mechanism), representing the fully open position of an SJM standard valve, as shown in Fig.~\ref{FlowCurve}.

\section{Governing equations}
Direct numerical simulations (DNS) are carried out in the present study. Under the assumption of an incompressible flow of blood, the following equations will govern the present flow dynamics

Continuity equation
\begin{equation} \label{eq:continuity}
   \bm{\nabla} \cdot \bm{u} = 0
\end{equation}

Momentum equation 
\begin{equation} \label{eq:momentum}
    \rho \left( \frac{\partial \bm{u}}{\partial t} + \bm{u} \cdot \bm{\nabla} \bm{u} \right) = - \bm{\nabla} p + \bm{\nabla} \cdot \bm{\tau} 
\end{equation}
In the above equations, $\bm{\nabla}$ is the gradient operator, $\bm{u}$ is the velocity vector, $t$ is the time, $p$ is the pressure, $\bm{\tau}$ is the extra-stress tensor, and $\rho$ is the density of blood (taken as $1060$ $kg/m^{3}$). The extra-stress tensor, $\bm{\tau}$, is evaluated as
\begin{equation} \label{eq:constitutive}
   \bm{\tau} = {\eta}\bm{\dot {\gamma}} 
\end{equation}
Where $\bm{\dot {\gamma}}$ is the shear-rate tensor and $\eta$ is the apparent shear viscosity of blood, which is evaluated by the three constitutive relations, namely, Newtonian (Eq.~\ref{eq:Newtonian}), power-law (Eq.~\ref{eq:power-law}) and Casson (Eq.~\ref{eq:Casson}), as follows 
\begin{equation} \label{eq:Newtonian}
   {\eta} = {\eta_0} 
\end{equation}

\begin{equation} \label{eq:power-law}
   {\eta} = k \left(\bm{\dot {\gamma}}\right)^{n-1}, \qquad {\eta_0} \leq {\eta} \leq {\eta_{\infty}} 
\end{equation}

\begin{equation} \label{eq:Casson}
   {\sqrt{\eta}} =  \sqrt{\tau_{0}/\bm{\dot {\gamma}}} + \sqrt{m}, \qquad {\eta_0} \leq {\eta} \leq {\eta_{\infty}} 
\end{equation}
Where $k$, $n$, $\eta_0$, $\eta_{\infty}$, $\tau_{0}$, and $m$ denote the fluid consistency coefficient, flow behavior index, zero-shear viscosity, infinite or high-shear viscosity, threshold or yield stress, and consistency index, respectively. As mentioned earlier, the values of all these parameters are obtained by fitting the experimental rheological response of real and whole blood conducted by Valant et al.~\citep{Valant2011}) as shown in Fig.~\ref{ShearViscosity} and those are as follows: $k$ = $0.018$ $Pa-s^{0.708}$, $n$ = $0.708$, $\eta_{0}$ = $0.0548$ $Pa-s$, $\eta_{\infty}$ = $0.0035$ $Pa-s$, $\tau_{0}$ = $0.004$ $Pa$, and $m$ = $0.0043$ $Pa-s$.

\section{Computational details}
\subsection{Solution algorithm}\label{Solution algorithm}
In this study, we have employed the finite volume method (FVM) based open-source computational fluid dynamics (CFD) code OpenFOAM (version 7)~\citep{openfoam} to solve all the governing equations numerically. Specifically, we have utilized the \textit{pisoFoam} solver to conduct simulations for various blood rheological models, including Newtonian, power-law, and Casson models. We have opted for the second-order Gauss linear scheme to discretize the momentum equation's advective terms. Time derivative terms have been discretized using the Euler scheme, while diffusion terms appearing in the momentum equation were handled with the second-order accurate Gauss linear corrected interpolation scheme. For solving the linear system of pressure fields, we have employed the Generalized Geometric-Algebraic MultiGrid (GAMG) solver coupled with the Gauss-Seidel smoother, chosen for its efficiency in information transport across the solution domain. Velocity fields were solved using the smooth solver with a symmetric type Gauss-Seidel preconditioner, capable of both forward and reverse sweeps. Pressure-velocity coupling was accomplished using the Pressure Implicit with Splitting of Operators (PISO) algorithm~\citep{PISO2001}, known for its improved efficiency over the Semi-Implicit Method for Pressure Linked Equations (SIMPLE) algorithm, particularly in unsteady problems. A relative tolerance level of $10^{-6}$ was set for both velocity and pressure. Simulations were conducted for three cardiac cycles for pulsatile flow (up to $t = 2580$ ms), whereas for steady flow, those were conducted up to $t = 5000$ ms using 200-640 processors in parallel. The simulations required approximately 60-144 hours of run time, equivalent to approximately 14,000-57,000 computational resource hours, on a supercomputer equipped with Intel Xeon Platinum 8268, 2.9 GHz processors, and 960 GB of memory per node.
Finally, to complete the problem setup, the following set of boundary conditions has been imposed.

\subsection{Boundary conditions}
\textbf{At the inlet boundary}: The fully developed flow condition was used for the steady inflow condition, whereas the prescribed flow rate (as shown in Fig.~\ref{FlowCurve}) condition was used for the pulsatile flow condition. On the other hand, the pressure was specified with a zero-gradient value at this boundary. 

\textbf{At the outlet boundary}: A Neumann-type boundary condition was used for the velocity, whereas the pressure value of $100$ $mm$ $of$ $Hg$ was provided at this boundary, which is the average of systolic ($\approx$ $120$ $mm$ $of$ $Hg$) and diastolic ($\approx$ $80$ $mm$ $of$ $Hg$) pressure of a normal human being.   

\textbf{At the leaflets and walls}: The no-slip boundary condition was imposed for the velocity, whereas the zero-gradient for the pressure was used at these surfaces.

\subsection{Grid and time-step size independence studies}\label{Grid and time-step size}
In addition, we have conducted systematic grid and time-step size independence studies to ensure the robustness and reliability of our present numerical simulations. These studies involved varying the grid resolution and time-step size while keeping other simulation parameters constant. Such investigations are essential for validating the numerical approach and ensuring consistent results regardless of the chosen discretization schemes and tolerance levels. We have first constructed the computational domain and initial mesh structure using the \textit{blockMeshDict} subroutine available within OpenFOAM. Subsequently, we have utilized \textit{snappyHexMeshDict}, a script available with OpenFOAM, to refine the mesh, particularly in areas of anticipated high gradients, such as around the leaflets. The mesh structure employed in this study comprised polyhedrons, chosen for their reduced sensitivity to stretching near walls and leaflets, which led to improved mesh quality and enhanced numerical stability.

\begin{figure}
    \centering
    \includegraphics[trim=0cm 0cm 0cm 0cm,clip,width=8cm]{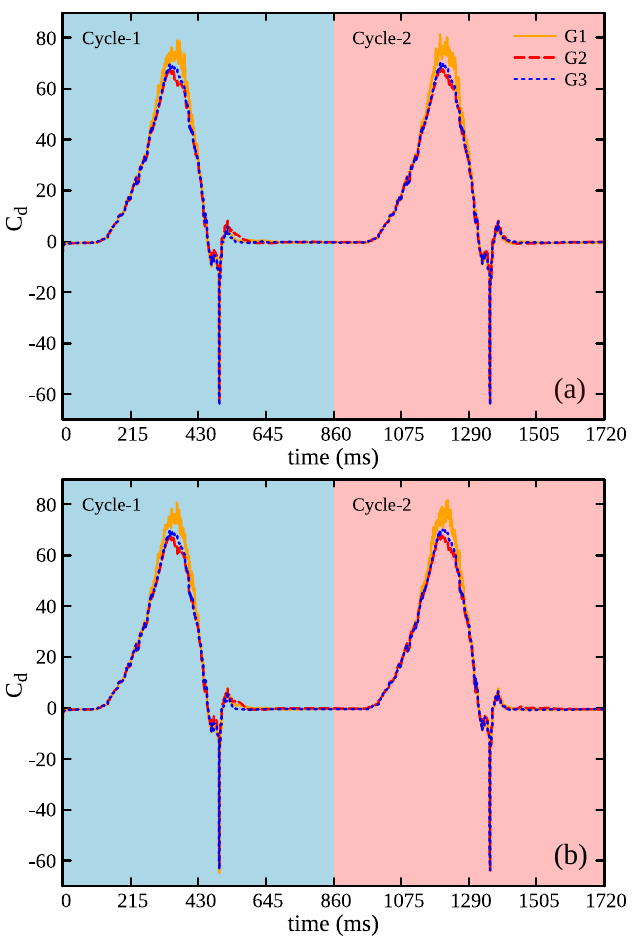}
    \caption{Temporal variation of non-dimensional drag forces ($C_d$) acting on the surface of the leaflet for three different grids (details given in Table~\ref{table:GridIndependence}) considered in the present study for the top (a) and bottom (b) leaflets.} 
    \label{GridCd}
\end{figure}

\begin{figure}
    \centering
    \includegraphics[trim=0cm 0cm 0cm 0cm,clip,width=8cm]{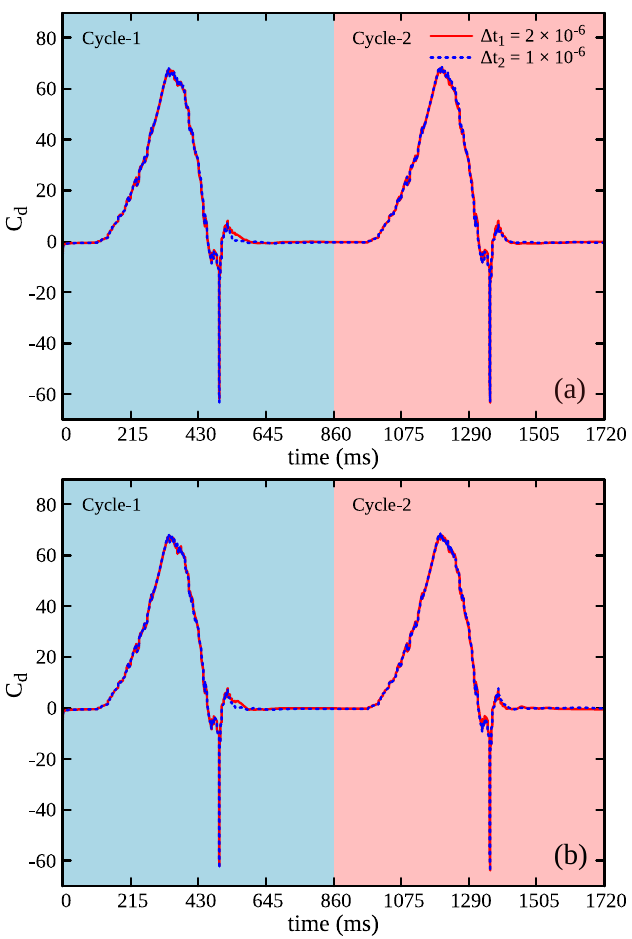}
    \caption{Temporal variation of non-dimensional drag forces ($C_d$) acting on the surface of the leaflet for two different time-step sizes (details given in Table~\ref{table:TimeStepSize}) considered in the present study for top (a) and bottom (b) leaflets.} 
    \label{TSCd}
\end{figure}

For the grid independence study, we have created three distinct grids, namely, G1, G2, and G3, with increasing numbers of cells, as detailed in Table~\ref{table:GridIndependence}. Simulations were conducted for blood flow under pulsatile conditions, corresponding to two cardiac cycles (i.e., $t = 1720$ ms), employing the power-law rheological model. Figure~\ref{GridCd} illustrates the temporal variation of non-dimensional drag forces (represented as drag coefficient, $C_{d}$) acting on the leaflet surfaces for the three grid configurations. The corresponding time-averaged $(<C_d>)$ and maximum $(C_{d,max})$ values of the drag coefficient are summarized in Table~\ref{table:GridIndependence}. Analysis of these results indicates that grid G2, comprising approximately 4.25 million cells, is adequate for the present study, with relative errors in time-averaged and maximum drag coefficient values of approximately $0.5223\%$ and $2.8789\%$ for the top leaflet, and $0.1405\%$ and $2.8773\%$ for the bottom leaflet, respectively, compared to grid G3. Furthermore, Figure~\ref{TSCd} and Table~\ref{table:TimeStepSize} compare results obtained with two different time-step sizes using grid G2 to evaluate the influence of time-step size selection. These comparisons demonstrate negligible differences as the time-step size decreases from $\Delta t_{1} = 2 \times 10^{-6}$ to $ \Delta t_{2} = 1 \times 10^{-6}$, with relative errors of less than 1\% across both cardiac cycles. Furthermore, to ensure numerical stability, we have monitored the Courant number ($Co$), defined as $Co = \frac{u \Delta t}{\Delta x}$, to satisfy the CFL (Courant-Friedrichs-Lewy) condition~\citep{courant1967}. Here, $\Delta t$ represents the time-step size, and $\frac{\Delta x}{u}$ denotes the characteristic convective time scale. Throughout our simulations, we have maintained $Co_{max} \le 0.9$ with $\Delta t \le 2 \times 10^{-6}$ s, ensuring stability in the numerical solution.

\begin{table*}
  \caption{Details of the grid independence study performed in the present study with the shear-thinning power-law fluid model of blood under pulsatile flow conditions. Note that here, all the parameter values are obtained after simulating two cardiac cycles, and $<.>$ denotes the time-averaged value of a parameter.}
  \begin{center}
  \begin{ruledtabular}
  \begin{tabular}{ccccc}
    & & {Grid 1 (G1)} & \textbf{Grid 2 (G2)} & Grid 3 (G3) \\[5pt]
    & {Hexahedra} & 2,524,198 & \textbf{3,598,868} & 4,883,516\\[2pt]
    \multirow{5}{*}{\rotatebox[origin=c]{90}{Polyhedra}} & {6 faces} & 88,538 & \textbf{119,434} & 157,826\\
    & {9 faces} & 292,128 & \textbf{402,490} & 525,104\\
    & {12 faces} & 61,480 & \textbf{84,768} & 112,234\\
    & {15 faces} & 32,544 & \textbf{45,854} & 61,724\\
    & {18 faces} & 242 & \textbf{352} & 348\\[2pt]
    & {Total number of cells} & 2,999,130 & \textbf{4,251,766} & 5,740,752\\[2pt]
    & {Minimum cell volume ($m^3$)} & $3.87 \times 10^{-16}$ & \textbf{$2.50 \times 10^{-16}$} & $1.37 \times 10^{-16}$\\[2pt]
    & {Maximum cell volume ($m^3$)} & $8.40 \times 10^{-9}$ & \textbf{$4.51 \times 10^{-9}$} & $2.70 \times 10^{-9}$\\[2pt]
    & {$<C_d>~$ (Top leaflet)} & 13.8472 & \textbf{12.7135} & 12.7799\\[2pt]
    & {\% Error~~} & $-$ & 8.1872 & 0.5223\\[2pt]    
    & {$C_{d, max}~$ (Top leaflet)} & 81.5010 & \textbf{68.1834} & 70.1463\\[2pt]    
    & {\% Error~~} & $-$ & 16.3404 & 2.8789\\[2pt]    
    & {$<C_d>~$ (Bottom leaflet)} & 13.8726 & \textbf{12.7360} & 12.7539\\[2pt]
    & {\% Error~~} & $-$ & 8.1931 & 0.1405\\[2pt]    
    & {$C_{d, max}~$ (Bottom leaflet)} & 81.6807 & \textbf{68.1994} & 70.1617\\[2pt]    
    & {\% Error~~} & $-$ & 16.5049 & 2.8773\\[2pt]    
  \end{tabular}
  \end{ruledtabular}  
  \label{table:GridIndependence}
  \end{center}
\end{table*}

\begin{table}
  \caption{Details of the time-step size convergence study performed in the present study with the shear-thinning power-law fluid model of blood under pulsatile flow conditions.}
  \begin{center}
  \begin{ruledtabular}
  \begin{tabular}{lcccc}
    & & \textbf{step size-1} (\bm{$\Delta t_{1}$}) & step size-2 ($\Delta t_{2}$) \\[5pt]
    \multicolumn{2}{c}{step size value (s)} & \bm{$2 \times 10^{-6}$} & {$1 \times 10^{-6}$}\\[2pt]
    \multicolumn{2}{c}{$<C_d>~$ (Top leaflet)} & \textbf{12.7135} & 12.6151\\[2pt]
    \multicolumn{2}{c}{\% Error~~} & $-$ & 0.7740\\[2pt]    
    \multicolumn{2}{c}{$C_{d, max}~$ (Top leaflet)} & \textbf{68.1834} & 68.5386\\[2pt]    
    \multicolumn{2}{c}{\% Error~~} & $-$ & 0.5209\\[2pt]    
    \multicolumn{2}{c}{$<C_d>~$ (Bottom leaflet)} & \textbf{12.7360} & 12.6564\\[2pt]
    \multicolumn{2}{c}{\% Error~~} & $-$ & 0.6250\\[2pt]    
    \multicolumn{2}{c}{$C_{d, max}~$ (Bottom leaflet)} & \textbf{68.1994} & 68.6880\\[2pt]    
    \multicolumn{2}{c}{\% Error~~} & $-$ & 0.7164\\[2pt]    
  \end{tabular}
  \end{ruledtabular}
  \label{table:TimeStepSize}
  \end{center}
\end{table}

\subsection{Code validation}\label{Code validation}

\begin{figure}
    \centering
    \includegraphics[trim=0cm 0cm 0cm 0cm,clip,width=8cm]{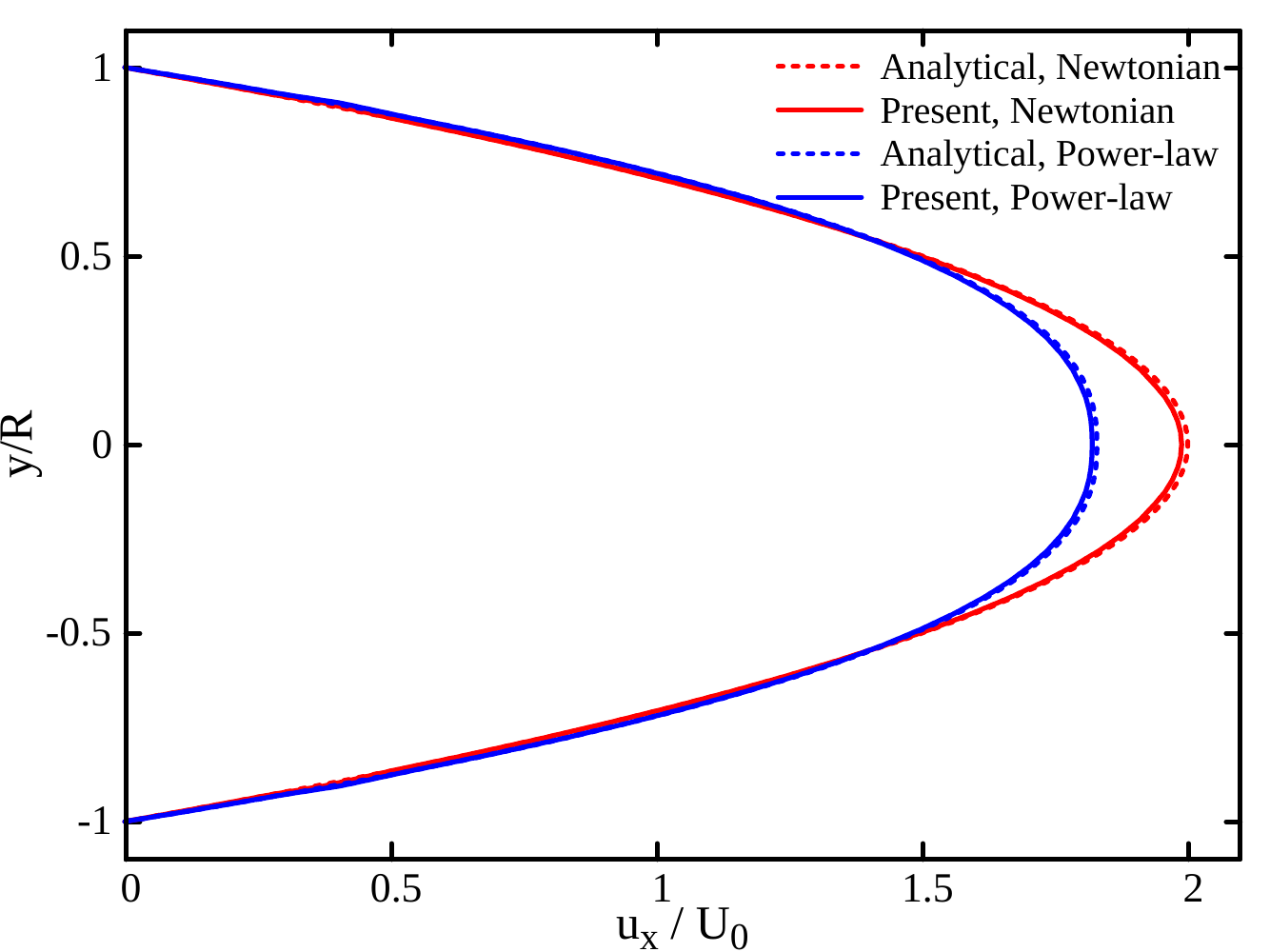}
    \caption{Comparison of the non-dimensional axial velocity profile obtained from the present implemented numerical code with that of the analytical solution for the flow in a circular pipe.} 
    \label{ValidationPipe}
\end{figure}

\begin{figure*}
    \centering
    \includegraphics[trim=0cm 0cm 0cm 0cm,clip,width=16cm]{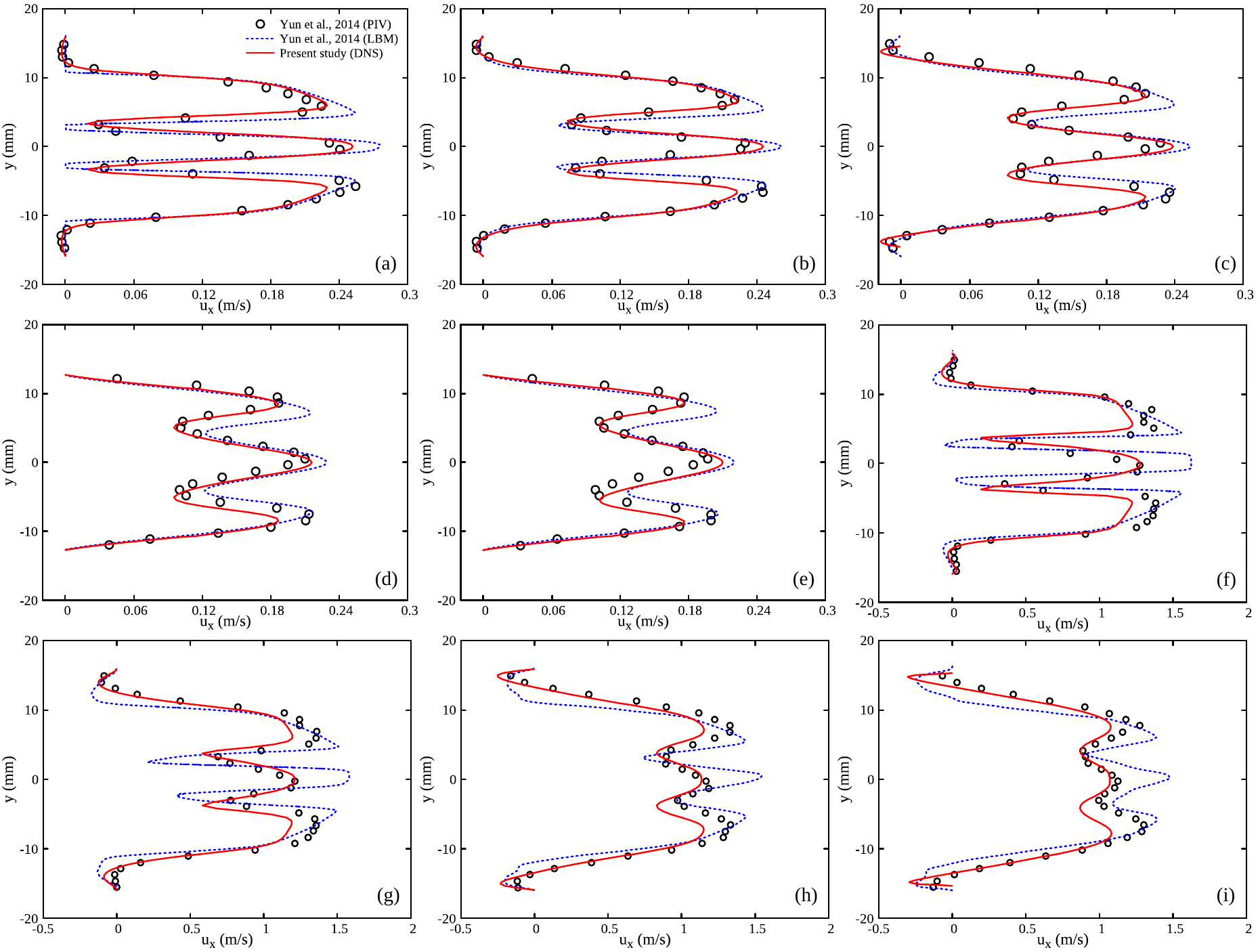}
    \caption{Validation of the present numerical setup for steady inflow at $Re = 750$ ((a)-(e)) and $Re = 5000$ ((f)-(i)) with that of experimental and numerical results of Yun et al.~\citep{Yun2014}. Here, the streamwise velocity component $(u_{x})$ is measured along the spanwise direction at different axial locations, namely, $x = 2.2$ (a), $x = 10.3$ (b), $x = 20.1$ (c), $x = 40.7$ (d), $x = 50.6$ (e), $x = 3.1$ (f), $x = 7.5$ (g), $x = 13.8$ (h), and $x = 18.3$ (i), where `x' is the distance measured in `mm' from the leaflet tip.} 
    \label{Validation1}
\end{figure*}

To ensure the accuracy and reliability of our numerical setup, we have conducted thorough code validation on benchmark problems before delving into the detailed analysis of our present new results. A key validation involved comparing the variation of the streamwise velocity component in a straight three-dimensional pipe between our numerical predictions and analytical solutions for both Newtonian and power-law fluid models. The analytical expressions for the axial velocity component are well-established, following the forms: $u_x = 2 U_0 \left(1-\left(\frac{2r}{D_0}\right)^2\right)$ for Newtonian fluids and $u_x = U_0 \left(\frac{3n+1}{n+1}\right)\left(1-\left(\frac{2r}{D_0}\right)^\frac{n+1}{n}\right)$ for power-law fluids~\citep{Chhabra2011}. Figure~\ref{ValidationPipe} illustrates this comparison, demonstrating excellent agreement between our numerical results and analytical predictions.

Moreover, we have conducted a comprehensive comparison with experimental and numerical studies that employed a similar computational setup, assuming blood is a simple Newtonian fluid. For instance, Yun et al.~\citep{Yun2014} conducted extensive \textit{in vitro} experiments and \textit{in silico} numerical simulations to analyze flow dynamics past a St. Jude Medical (SJM) Regent$^{TM}$ mechanical heart valve. Their experiments utilized digital particle image velocimetry (DPIV), while simulations were performed using the entropic lattice-Boltzmann method (LBM). Figure~\ref{Validation1} compares streamwise velocity variations along the spanwise direction at different locations downstream of the valve leaflet between our numerical predictions and the experimental and numerical results of Yun et al. at Reynolds numbers 750 and 5000 for steady inflow conditions. Overall, a good correspondence was observed, particularly between our direct numerical simulation (DNS) results and Yun et al.'s experimental findings. Additionally, further validations were also conducted for blood flow in our previous study, as detailed elsewhere~\citep{Chauhan2021}. Collectively, these validation efforts instill confidence in our ability to present and discuss the novel findings of this study, as outlined in the subsequent section.

\section{Results and discussion}\label{Results and discussion}
After solving the aforementioned governing equations with the above-stated boundary conditions, we have obtained the results in terms of velocity, pressure, and stress fields. These results are then further post-processed to calculate the parameters of clinical importance, such as wall shear stress ($WSS$) acting on the solid walls, non-dimensional drag forces ($C_d$) acting on the surface of the two valve leaflets, and the blood damage ($BD$). The mathematical expressions for evaluating $WSS$, $C_d$, and $BD$ are as follows:
\begin{equation}
   WSS = \bm{\tau} \cdot \bm{n_i} 
\end{equation}

\begin{equation}
   C_d = \frac{\bm{F_d}}{\frac{1}{2} \rho U_0^2 \bm{A_p}} = \frac{2}{\rho U_0^2 \bm{A_p}}\:\int_{S}\left(-p \bm{\delta} + \bm{\tau}\right) \cdot \bm{n_o} \cdot \bm{i} \:dS
\end{equation}

\begin{equation}
   BD = (\overline{D_{I}})^{0.785} 
\end{equation}
where $\bm{n_i}$ is the patch normal vector drawn into the domain, $U_0$ is the average velocity (for pulsatile flow, it is $\approx$ 0.148 $m/s$, and for steady inflow, its value depends on the Reynolds number), $\bm{A_p}$ is the projected area of the leaflet in the direction of flow, $\bm{F_d}$ is the dimensional drag forces acting on the leaflet, $\bm{\delta}$ is the Kronecker delta, $\bm{n_o}$ is the outward unit normal vector drawn on the leaflet surface, $\bm{i}$ is the unit vector in the x-direction, $S$ is the leaflet surface area, and $\overline{D_{I}}$ is the average linear damage calculated using the expression~\cite{garon2004fast}:

\begin{equation}
   \overline{D_{I}} = \frac{1}{Q}\:\int_{V}\bm{\sigma} \:dV 
\end{equation}

\begin{equation}
   \sigma = (3.62 \times 10^{-7})^{1/0.785} (\bm{\tau_{vm}})^{2.416/0.785} 
\end{equation}
where $Q$ is the volumetric flow rate, $\sigma$ is the rate of hemolysis production per unit time, $V$ is the volume of the whole computational domain, and $\bm{\tau_{vm}}$ is the Von Mises criterion calculated as proposed by Garon and Farinas~\citep{garon2004fast}.

We have conducted direct numerical simulations (DNS) to gain deeper insights into the flow dynamics of blood through a bileaflet St. Jude Medical heart valve (SJMHV). Our simulations considered blood as both Newtonian and non-Newtonian fluids to facilitate qualitative and quantitative comparisons between the two cases. Although blood primarily exhibits shear-thinning behavior, it also manifests threshold or yield stress characteristics, as previously mentioned and discussed in Fig.~\ref{ShearViscosity}. In our study, we have characterized blood's shear rheology using the power-law model, while the Casson fluid model was employed to represent yield stress. Simulations have been conducted under both steady and pulsatile flow conditions. For steady flow, we have enforced fully developed flow conditions at the inlet section of the ventricular side chamber, as depicted in Fig.~\ref{Geometry}. Conversely, for pulsatile flow, we have applied prescribed flow rate conditions, as illustrated in Fig.~\ref{FlowCurve}, at the same boundary. In the case of steady flow, we have performed both instantaneous and time-averaged analyses. For pulsatile flow, only instantaneous analyses have been carried out, with results presented at three distinct time instances: mid-acceleration phase (T1), peak phase (T2), and mid-deceleration phase (T3) during the systolic condition, as outlined in Fig.~\ref{FlowCurve}. Our analyses have encompassed various parameters, including velocity magnitude contours, pressure drop across inlet and outlet sections, Reynolds stresses, turbulent kinetic energy (TKE), wall shear stress ($WSS$), drag coefficient ($C_d$), local non-Newtonian importance factor ($I_L$), among others. These results are comprehensively presented and discussed to provide a detailed understanding of the differences in the flow characteristics caused by blood rheology. 

\begin{figure*}
    \centering
    \includegraphics[trim=0cm 0cm 0cm 0cm,clip,width=16cm]{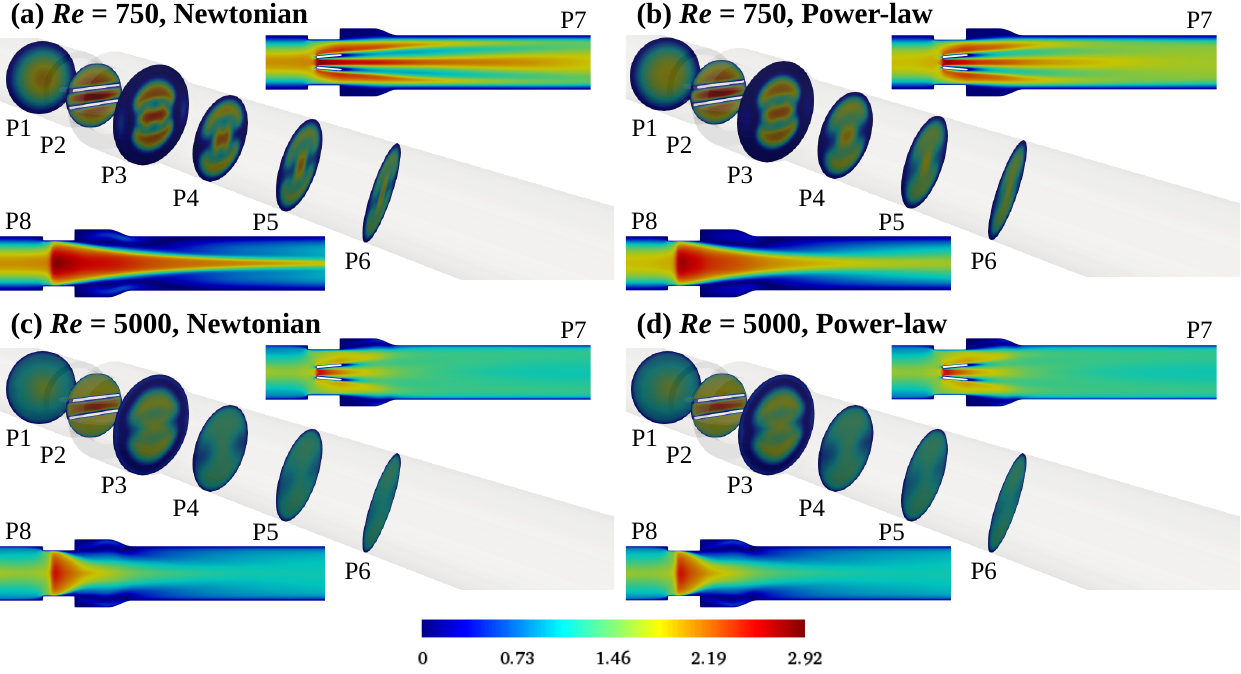}
    \caption{Non-dimensional velocity magnitude contours at eight different planes (P1-P8) for various Reynolds numbers predicted by Newtonian and non-Newtonian power-law fluid models of blood. Note that the non-dimensionalization for the velocity is performed using the average velocity imposed at the inlet plane for each case according to the value of $Re$, and the results shown here for $Re = 750$ are seen to be steady in nature and for $Re = 5000$ are time-averaged from $0 - 5000$ $ms$.} 
    \label{UmagSteady}
\end{figure*}

\begin{figure}
    \centering
    \includegraphics[trim=0cm 0cm 0cm 0cm,clip,width=8cm]{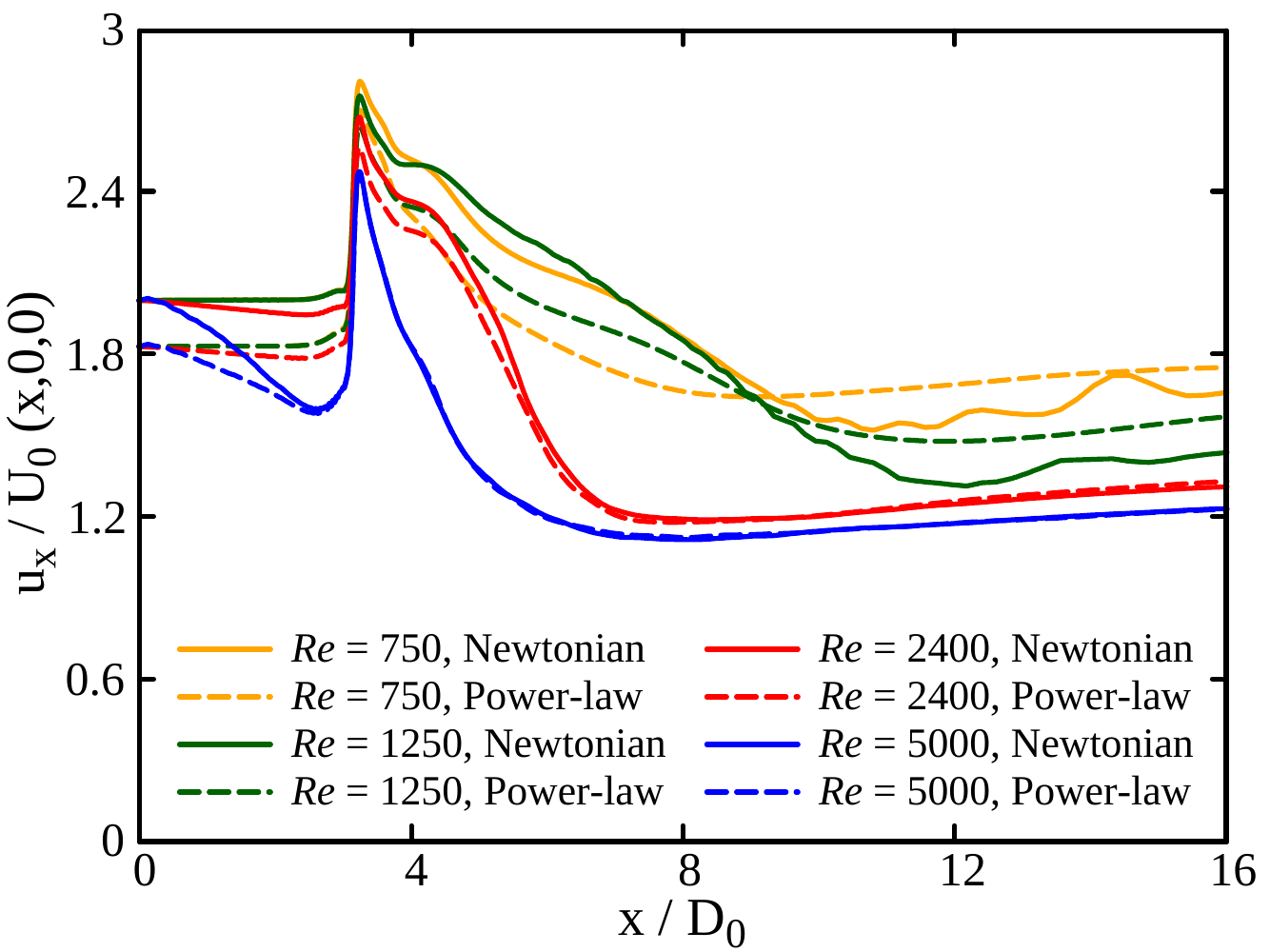}
    \caption{Variation of non-dimensional streamwise velocity component along the centerline for several Reynolds numbers predicted by Newtonian and non-Newtonian power-law fluid models of blood. Note that the results presented here are either steady or time-averaged, as mentioned in Table~\ref{table:SteadyfixedQ}.} 
    \label{UCenterline}
\end{figure}

\begin{figure*}
    \centering
    \includegraphics[trim=0cm 0cm 0cm 0cm,clip,width=16cm]{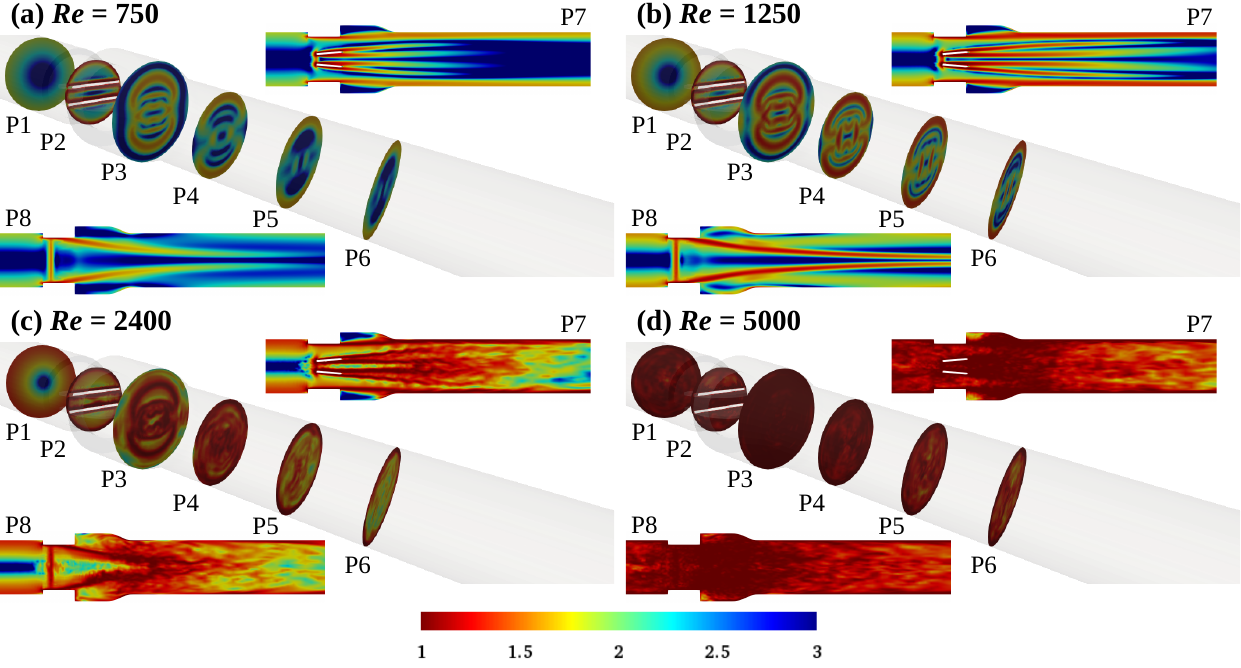}
    \caption{Instantaneous local non-Newtonian importance factor ($I_L=\eta/\eta_{\infty}$) at eight different planes (P1-P8) for various $Re$ values for power-law fluid model of blood, namely, (a) $Re = 750$, (b) $Re = 1250$, (c) $Re = 2400$, and (d) $Re = 5000$.} 
    \label{ILSteady}
\end{figure*}

\subsection{Steady inflow conditions}

Under steady inflow conditions, simulations have been carried out at four distinct Reynolds numbers, namely, 750, 1250, 2400, and 5000, wherein the Reynolds number is defined as $Re = \frac{4 \rho Q}{\pi D_{0} \eta_{\infty}}$. These four Reynolds numbers nearly correspond to the early beginning of the mid-acceleration phase, between the beginning and mid-acceleration phase, the mid-acceleration phase, and the peak phase during the systolic phase of our cardiac cycle, respectively~\cite{Yun2014}, as schematically shown in Fig.~\ref{FlowCurve}. At first, we present the variation of the non-dimensional velocity magnitude at the lowest $(Re = 750)$ and highest $(Re = 5000)$ Reynolds numbers encompassed in this study by considering blood both as Newtonian and non-Newtonian power-law fluids, Fig.~\ref{UmagSteady}. Eight different planes have been chosen to display the results: six planes (P1 to P6) normal to the x-direction with one situated upstream of the leaflet (P1) and the remaining situated downstream of the leaflet (P2-P6), one normal to the y-direction (P8), and the other normal to the z-direction (P7). The flow remains steady at $Re = 750$, whereas it becomes unsteady at $Re = 5000$. Therefore, the time-averaged results (for a time of 5000 $ms$, which corresponds to more than five times the time corresponding to a full cardiac cycle) are presented in Fig.~\ref{UmagSteady} for the latter case. The velocity varies smoothly, with the maximum at the axis of the ventricular side chamber and the minimum at the wall upstream of the valve leaflet at $Re = 750$; see plane P1 in sub-Figs.~\ref{UmagSteady} (a) and (b). As the blood approaches plane P2 (situated in the valve chamber region), the velocity field becomes more uniform across the plane, and its magnitude also increases. This is because the cross-sectional area of the valve region is less compared to the ventricular side chamber region, and the velocity field has to be accelerated to obey the mass conservation principle. Furthermore, at this plane, three fluid jets are formed, namely, one central jet formed between the two valve leaflets and two lateral jets formed between the valve leaflet and the wall of the chamber. The formation of these jets is clearly visible in plane P7. As blood passes through plane P3 (situated in the sinus region), once again, the velocity field becomes highly non-uniform in nature due to the larger cross-sectional area of this region, and three regions of high-velocity magnitude are seen due to the formation of three fluid jets. As blood traverses more downstream to the aortic chamber region (region D) downstream of the valve, a mushroom-like velocity field develops, which extends gradually towards the chamber wall as the downstream distance increases. The intensity of the fluid jets decreases with the distance downstream of the valve, which is noticeable both in planes P7 and P8. Therefore, far downstream of the valve, one can expect the velocity field to be the same as observed upstream of the valve. 

At a relatively higher Reynolds number of 5000, the same trend is qualitatively observed in the time-averaged results as that seen at $Re = 750$. However, some obvious quantitative differences are present. For instance, the velocity field at this Reynolds number is more uniform and dispersed at any plane perpendicular to the x-direction than seen at $Re = 750$. The magnitude of the central jet and two lateral jets is less, and they end in the sinus region (region C) at this $Re = 5000$ irrespective of the blood rheology type, i.e., Newtonian or non-Newtonian power-law (P7, sub-fig.~\ref{UmagSteady}(c) \& (d)). Also, the merging of the central jet with the two lateral jets happens quickly at this Reynolds number, unlike the case of $\textit{Re} = 750$, wherein they merged far downstream of the two leaflets (P1-P6, sub-fig.~\ref{UmagSteady}(c) \& (d)). It happens due to the increase in the flow strength at this value of the Reynolds number, which leads to more fluid movement in the lateral direction, ultimately facilitating the quick merging of the jets. However, the thickness of the central jet is higher at this Reynolds number than $Re = 750$. All these can be clearly visible from the results presented in planes P7 and P8 in sub-Figs.~\ref{UmagSteady}(c) and (d). Furthermore, the mushroom-like velocity field downstream of the valve is not clearly developed at this Reynolds number. Regardless of the blood rheology, all these differences are seen between these two Reynolds numbers. This is because the flow becomes unsteady and turbulent-like at $Re = 5000$ in contrast to a steady flow at $Re = 750$, which will be discussed in detail later in this section.

The effect of blood non-Newtonian behaviors on the spatial variation of the velocity field is also apparent. For instance, the velocity magnitude of the central and lateral jets is higher for the Newtonian model than the shear-thinning power-law model, particularly in the vicinity of the valve leaflets. Furthermore, the central jet is seen to be more extended downstream of the valve for the Newtonian model than for the power-law one. All these differences can be more evident on planes P7 and P8. However, these differences are seen to be more pronounced at  $Re = 750$ than at $Re = 5000$. This is probably because, at the higher Reynolds number, the shear rate becomes so high that the apparent viscosity reaches the Newtonian plateau (see Fig.~\ref{ShearViscosity}), resulting in the difference between the two models being minimal. A more quantitative difference between the two models is presented in Fig.~\ref{UCenterline} wherein the non-dimensional streamwise velocity is plotted along the horizontal line passing through the origin of the geometry. The streamwise velocity is seen to be maximum at around $x/D_{0} = 4$, which is the region between the two leaflets. It then gradually decreases as one moves further downstream of the valve. The maximum velocity for the Newtonian model is higher than for the power-law one, and the difference between the two decreases as the Reynolds number increases. The same trend is also seen downstream of the valve. All in all, the difference between the two models is seen to be more noticeable downstream of the valve than upstream of it. This is because the detachment of the shear layers happens downstream of the valve leaflets, resulting in the generation of vortices wherein the fluid rheology would have a significant impact on the velocity field.      

To precisely assess the impact of blood's non-Newtonian rheological properties on hemodynamics, in the present study, we calculate the non-Newtonian importance factor ($I$) proposed by Ballyk et al.~\cite{ballyk1994}. This factor is defined as the ratio of the effective (or apparent) viscosity ($\eta_{eff}$) to the high-shear viscosity ($\eta_{\infty}$) of the fluid, expressed as $I = \frac{\eta_{eff}}{\eta_{\infty}}$. Subsequently, Johnston et al.~\cite{johnston2004} extended this concept to compute the local non-Newtonian importance factor ($I_{L}$) at specific points within the flow system. The value of $I_{L}$ equals one in the case of Newtonian flow, whereas deviations from this value indicate the influence of non-Newtonian flow characteristics of the blood. Figure~\ref{ILSteady} illustrates the instantaneous surface distribution of $I_{L}$ for the power-law model across various planes within the present cardiovascular system at different Reynolds numbers. At a low Reynolds number ($Re = 750$, sub-Fig.~\ref{ILSteady})(a), most regions, particularly central areas, exhibit $I_{L}$ values exceeding one, except for regions near the leaflet, valve housing wall, and around jets formed in the sinus region, where $I_{L}$ value remains around one. With increasing Reynolds numbers, regions with $I_{L}$ values near one expand radially and axially within the flow system. At the highest Reynolds number ($Re = 5000$, sub-Fig.~\ref{ILSteady}(d)), $I_{L}$ approximates one across the entire region, indicating diminished influence of blood's non-Newtonian rheology on hemodynamics. This suggests a more pronounced effect of non-Newtonian properties of blood at lower Reynolds numbers, which will gradually decrease with increasing Reynolds numbers. This will be reflected in the discussions later in this section on the variation of the clinically relevant surface-averaged parameters, such as pressure drop and wall shear stress, with the Reynolds number.  

\begin{figure}
    \centering
    \includegraphics[trim=0cm 0cm 0cm 0cm,clip,width=8cm]{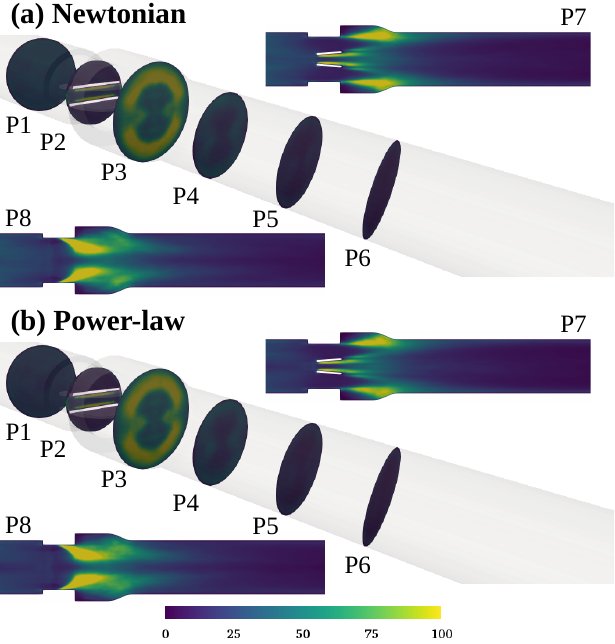}
    \caption{Surface variation of the Reynolds normal stress (RNS) at eight different planes (P1-P8) for $Re = 5000$ predicted by (a) Newtonian and (b) non-Newtonian power-law model of blood.} 
    \label{RNS}
\end{figure}

\begin{figure}
    \centering
    \includegraphics[trim=0cm 0cm 0cm 0cm,clip,width=8cm]{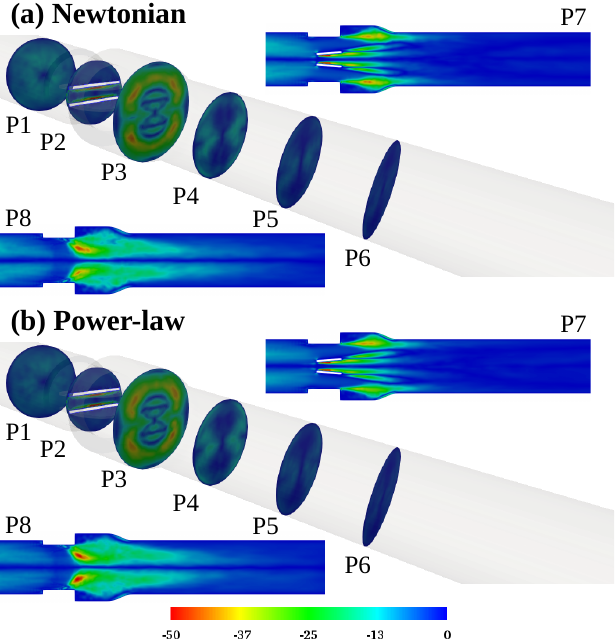}
    \caption{Surface variation of the Reynolds shear stress (RSS) at eight different planes (P1-P8) for $Re = 5000$ predicted by (a) Newtonian and (b) non-Newtonian power-law model of blood.} 
    \label{RSS}
\end{figure}

The flow field exhibits a steady and laminar behavior at a Reynolds number of $Re = 750$, nearly corresponding to the initial stage of the mid-acceleration phase in the cardiac cycle. However, as the cardiac cycle progresses, the flow undergoes a transition from steady and laminar to unsteady and turbulent conditions. This transition is particularly pronounced at the peak phase of the cardiac cycle, where the Reynolds number peaks at nearly 5000. Consequently, we investigate the surface distribution of Reynolds normal (RNS) and shear (RSS) stresses for both Newtonian and non-Newtonian power-law blood models at $Re = 5000$, as depicted in Figs.~\ref{RNS} and~\ref{RSS}, respectively. Reynolds stresses, also often referred to as turbulent stresses, quantify the intensity of velocity fluctuations associated with turbulent eddies in the flow. Normal stresses represent fluctuations occurring perpendicular to each other, while shear stresses represent fluctuations parallel to each other. In the context of cardiovascular flow, studying these turbulent stresses is crucial due to their potential to cause damage to blood cells~\cite{jones1995relationship, yen2014effect,sallam1984human}. Analysis of the Reynolds stresses reveals higher values of both normal and shear stresses in the region between the two leaflets and at the edges of lateral jets formed in the sinus region, regardless of the blood rheological model employed. This suggests that the intensity of velocity fluctuations is highest in these regions, indicating a higher risk of blood cell damage. Interestingly, the surface distribution patterns of turbulent stresses are similar for both Newtonian and power-law models, although the magnitude is slightly higher for the Newtonian model. This discrepancy is attributed to the minimal difference in flow dynamics between the two models at the high Reynolds numbers associated with the peak phase of the cardiac cycle. This observation is consistent with the distribution of the local non-Newtonian factor, which remains close to one throughout the region, indicating minimal variation between the two rheological models.

\begin{figure}
    \centering
    \includegraphics[trim=0cm 0cm 0cm 0cm,clip,width=8cm]{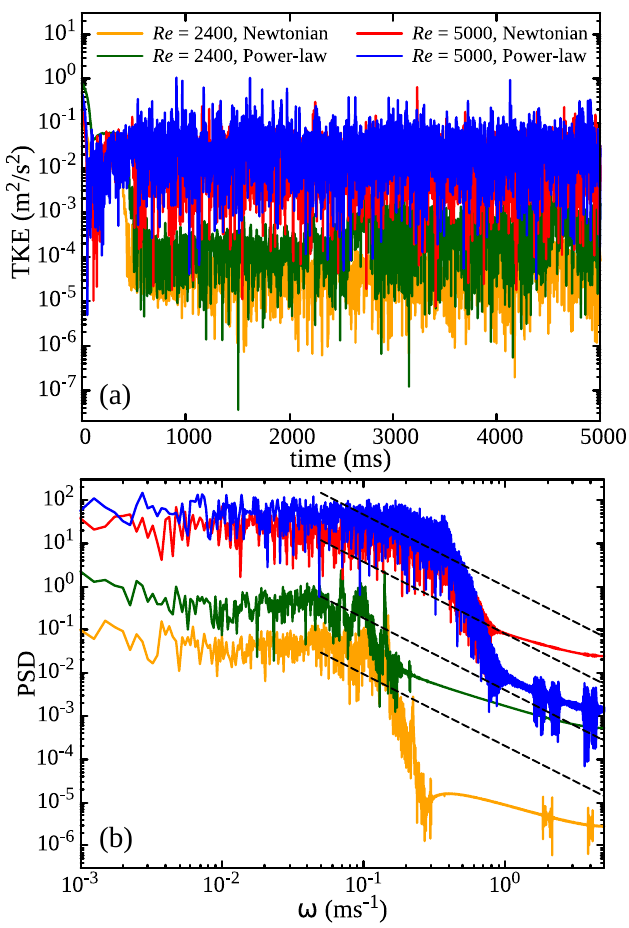}
    \caption{Time history of turbulent kinetic energy (TKE) at a probe situated near the tip of the two leaflets (a) and the corresponding power spectral density plot (b). The dashed black lines in the second sub-figure denote the lines with a constant slope of $-5/3$ corresponding to the Kolmogorov energy scaling law of turbulence.} 
    \label{TKE}
\end{figure}

\begin{figure}
    \centering
    \includegraphics[trim=0cm 0cm 0cm 0cm,clip,width=8cm]{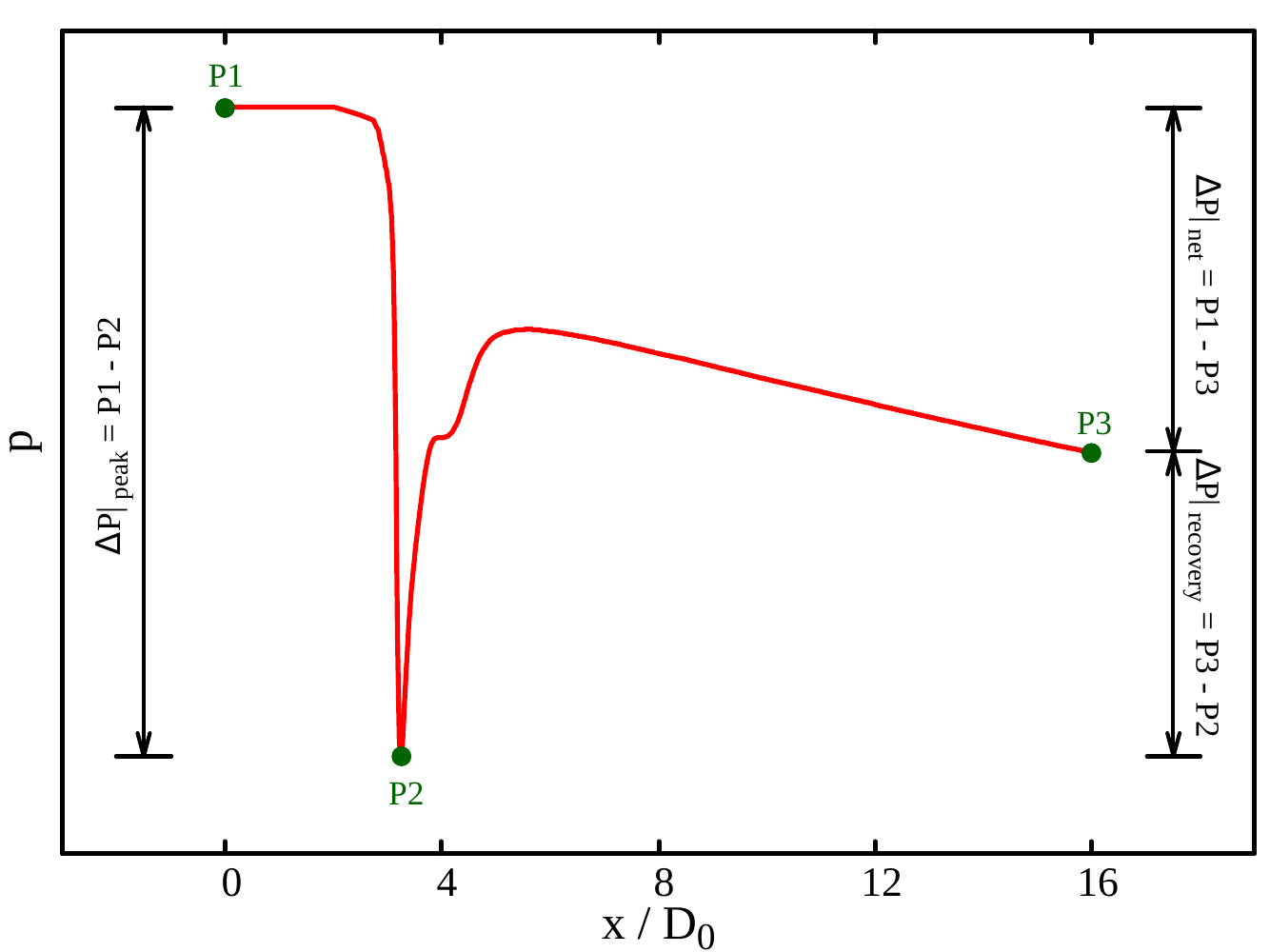}
    \caption{Schematic of the pressure variation along the center line and the various locations chosen to show different types of pressure drop calculated in Table~\ref{table:SteadyfixedQ}.} 
    \label{pVariation}
\end{figure}

\begin{figure}
    \centering
    \includegraphics[trim=0cm 0cm 0cm 0cm,clip,width=6cm]{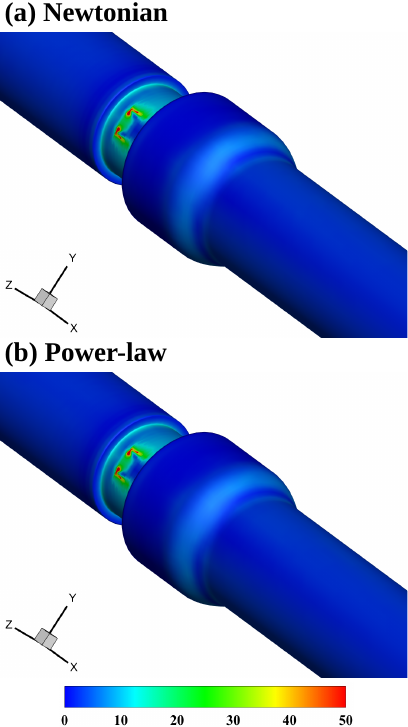}
    \caption{Time-averaged wall shear stress (WSS) magnitude at $Re = 5000$ for (a) Newtonian and (b) non-Newtonian power-law fluid models. The values shown in the legend have units of '$Pa$'.} 
    \label{WSSRe5000}
\end{figure}

Figure~\ref{TKE} illustrates the temporal evolution of turbulent kinetic energy (TKE) per unit mass of fluid at a specific probe location positioned between the two leaflets on the front side, along with its corresponding power spectral density (PSD) plot. This analysis is conducted for two different blood rheological models at Reynolds numbers 2400 and 5000. TKE is computed as half the sum of the variances of the fluctuating velocity components, defined as $\frac{1}{2}\left( \overline{(u_{x}')^{2}} + \overline{(u_{y}')^{2}} + \overline{(u_{z}')^{2}}\right)$, where $u'$ represents the fluctuating velocity component ($u' = u - \overline{u}$), and $u$ and $\overline{u}$ denote instantaneous and average velocity components, respectively. The temporal variation of TKE demonstrates an increase with Reynolds number, indicating higher turbulence intensity, as depicted in sub-Fig.~\ref{TKE}(a). At $Re = 2400$, the TKE magnitude is notably greater for the non-Newtonian power-law model than the Newtonian model, though this difference diminishes at $Re = 5000$. This trend is also reflected in the power spectrum of TKE fluctuations shown in sub-Fig.~\ref{TKE}(b). The black dashed lines in this plot represent the slope of -5/3, corresponding to the Kolmogorov inertial subrange of the energy cascade. Notably, the observed slope deviates from -5/3 at both Reynolds numbers, irrespective of the blood rheological model, indicating the presence of non-Kolmogorov turbulent flow conditions within the heart valve. Recent studies~\cite{saqr2020physiologic,saqr2022non,saqr2022non2} have highlighted the prevalence of non-Kolmogorov turbulent flow in physiological blood flow across various geometries, attributed to the highly inhomogeneous and anisotropic nature of turbulence induced by the presence of solid valve leaflets.               

Finally, Table~\ref{table:SteadyfixedQ} represents various post-processed parameters, including some surface and time-averaged (in the case of unsteady flow) parameters that provide more quantitative analysis of the effect of blood non-Newtonian behaviors on the hemodynamics past the present mechanical heart valve. First of all, three different pressure difference measurements are calculated in this study, namely, $\Delta P|_{peak}$,  $\Delta P|_{recovery}$, and  $\Delta P|_{net}$. Their definitions are schematically shown in Fig.~\ref{pVariation}.  To calculate all these parameters, the values of the pressure field are first extracted along a horizontal line passing through the center of the region between the two leaflets. This horizontal line's starting (P1) and ending (P3) points are situated at the inlet and outlet of the computational domain. The pressure will be higher upstream of the valve leaflet at point P1. It will be minimal in the region between the two leaflets at point P2 due to the increased velocity magnitude in this region. However, some recovery will happen at point P3 downstream of the valve leaflets. Irrespective of the blood rheological model, the values of all three parameters related to the pressure difference increase as the Reynolds number increases. On the other hand, the values of $\Delta P|_{peak}$ and $\Delta P|_{net}$ are always higher in the case of the power-law rheological model of blood than the Newtonian model. In contrast, a reverse trend is seen for $\Delta P|_{recovery}$ at all Reynolds numbers. This suggests that the recovery of pressure downstream of the valve leaflets is more in the case of the Newtonian model than the non-Newtonian power-law model. However, the difference between the values of the two models diminishes as the Reynolds number increases. For instance, at $Re = 750$ the differences between the values of $\Delta P|_{peak}$,  $\Delta P|_{recovery}$, and  $\Delta P|_{net}$ are around 10\%, 100\%, and 46\%, respectively, whereas the corresponding differences at $Re = 5000$ are 0.86\%, 3.27\%, and 2.80\%, respectively. This highlights that blood rheology will have a greater influence on the parameters of clinical importance at lower Reynolds numbers. This is expected as it was also evident in the distribution of the local non-Newtonian importance factor $I_{L}$ presented in Fig.~\ref{ILSteady}, whose value approached nearly one in the entire region of the flow system, suggesting a minimal influence of blood non-Newtonian behaviors on the hemodynamics.

Next, the hydrodynamic drag forces acting on the valve leaflets are examined for two blood rheological models, which is crucial for optimizing the design and performance of mechanical heart valves. Contrary to the pressure difference, drag forces decrease with increasing Reynolds number, regardless of the blood rheological model. This decline is attributed to the reduction in the frictional component of drag forces. Notably, the difference between the two models is more pronounced at lower Reynolds numbers, gradually diminishing as the Reynolds number increases. For example, at $Re = 750$, the difference in drag forces is approximately 4\%, decreasing to less than 1\% at $Re = 5000$. Additionally, the maximum and surface-averaged (and time-averaged for unsteady flow) values of wall shear stress (WSS) are analyzed for both blood rheological models. WSS is a fundamental biomechanical factor in hemodynamics, regulating vascular function, pathophysiology, and therapeutic interventions. Understanding its role is essential for elucidating cardiovascular diseases and developing targeted therapies for their prevention and treatment~\cite{cecchi2011role,malek1999hemodynamic,davies2009hemodynamic,boussel2008aneurysm}. The time-averaged surface distribution of WSS, depicted in Fig.~\ref{WSSRe5000}, reveals higher values around the valve leaflets, particularly in the hinge region, regardless of the blood rheological model. Significant values are also observed on the valve housing region walls (section B in Fig.~\ref{Geometry}), indicating an increased risk of blood cell damage in these regions~\cite{leverett1972red}. Furthermore, the maximum and surface-averaged WSS values are consistently higher in the non-Newtonian power-law model compared to the Newtonian model. Similar to drag forces, the difference between the two models is more pronounced at lower Reynolds numbers, gradually diminishing as the Reynolds number increases. For instance, differences are approximately 67\% and 0.9\% at $Re = 750$ and 5000, respectively, for surface-averaged WSS. In summary, differences in various post-processed parameters between the two blood rheological models are evident at lower Reynolds numbers, corresponding to the beginning of the mid-acceleration phase of the cardiac cycle, and gradually diminish as the flow progresses to its peak phase. Therefore, these findings highlight the importance of considering blood non-Newtonian behaviors in the design and evaluation of cardiovascular devices and treatments.

\begin{table*}[hbt!]
  \caption{Values of various post-processed parameters for Newtonian and non-Newtonian power-law models of blood obtained at different Reynolds numbers.}
  \begin{center}
  \begin{threeparttable}
  \begin{ruledtabular}
  \begin{tabular}{ccccccc}
    \textbf{Parameter} & \textbf{Fluid Model} & \bm{$Re = 750$}  & \bm{$Re = 1250$}  & \bm{$Re = 2400$}  & \bm{$Re = 5000$}  \\[5pt]
    \multirow{2}{*}{$\Delta P|_{peak}$ ($mmHg$)} & {Newtonian} & $0.1590$\tnote{*} & $0.3975$\tnote{*} & $1.4310$\tnote{\#} & $6.4395$\tnote{\#}\\
    & {Power-law} & $0.1749$\tnote{*} & $0.4055$\tnote{*} & $1.4469$\tnote{\#} & $6.4952$\tnote{\#} \\[2pt]
    \multirow{2}{*}{$\Delta P|_{recovery}$ ($mmHg$)} & {Newtonian} & $0.0398$\tnote{*} & $0.1749$\tnote{*} & $0.3896$\tnote{\#} & $3.0290$\tnote{\#}\\
    & {Power-law} & $0$\tnote{*} & $0.1193$\tnote{*} & $0.3339$\tnote{\#} & $2.9298$\tnote{\#} \\[2pt]
    \multirow{2}{*}{$\Delta P|_{net}$ ($mmHg$)} & {Newtonian} & $0.1193$\tnote{*} & $0.2226$\tnote{*} & $1.0415$\tnote{\#} & $3.4106$\tnote{\#}\\
    & {Power-law} & $0.1749$\tnote{*} & $0.2862$\tnote{*} & $1.1130$\tnote{\#} & $3.5060$\tnote{\#} \\[2pt]
    \multirow{2}{*}{$C_d$ (Top leaflet)} & {Newtonian} & $4.6749$\tnote{*} & $3.5792$\tnote{*} & $2.7348$\tnote{\#} & $2.4063$\tnote{\#} \\
    & {Power-law} & $4.8745$\tnote{*} & $3.6060$\tnote{*} & $2.7057$\tnote{\#} & $2.3844$\tnote{\#} \\[2pt]
    \multirow{2}{*}{$C_d$ (Bottom leaflet)} & {Newtonian} & $4.6749$\tnote{*} & $3.5792$\tnote{*} & $2.7397$\tnote{\#} & $2.4039$\tnote{\#} \\
    & {Power-law} & $4.8745$\tnote{*} & $3.6060$\tnote{*} & $2.7071$\tnote{\#} & $2.3970$\tnote{\#} \\[2pt]
    \multirow{2}{*}{$|WSS|^{max}_{walls}$ ($Pa$)} & {Newtonian} & $2.7811$\tnote{*} & $6.4044$\tnote{*} & $19.9064$\tnote{\#} & $80.2681$\tnote{\#} \\
    & {Power-law} & $2.8711$\tnote{*} & $6.7006$\tnote{*} & $20.3392$\tnote{\#} & $79.2888$\tnote{\#} \\[2pt]
    \multirow{2}{*}{$|WSS|^{avg}_{walls}$ ($Pa$)} & {Newtonian} & $0.1419$\tnote{*} & $0.2647$\tnote{*} & $0.7478$\tnote{\#} & $2.4714$\tnote{\#} \\
    & {Power-law} & $0.2380$\tnote{*} & $0.3699$\tnote{*} & $0.8247$\tnote{\#} & $2.4484$\tnote{\#} \\[2pt]
  \end{tabular}
  \begin{tablenotes}
   \item[*] steady
   \item[\#] time-averaged
  \end{tablenotes}  
  \end{ruledtabular}
  \label{table:SteadyfixedQ}
  \end{threeparttable}
  \end{center}
\end{table*}

\subsection{Pulsatile flow conditions}
 \begin{figure}
    \centering
    \includegraphics[trim=0cm 0cm 0cm 0cm,clip,width=8cm]{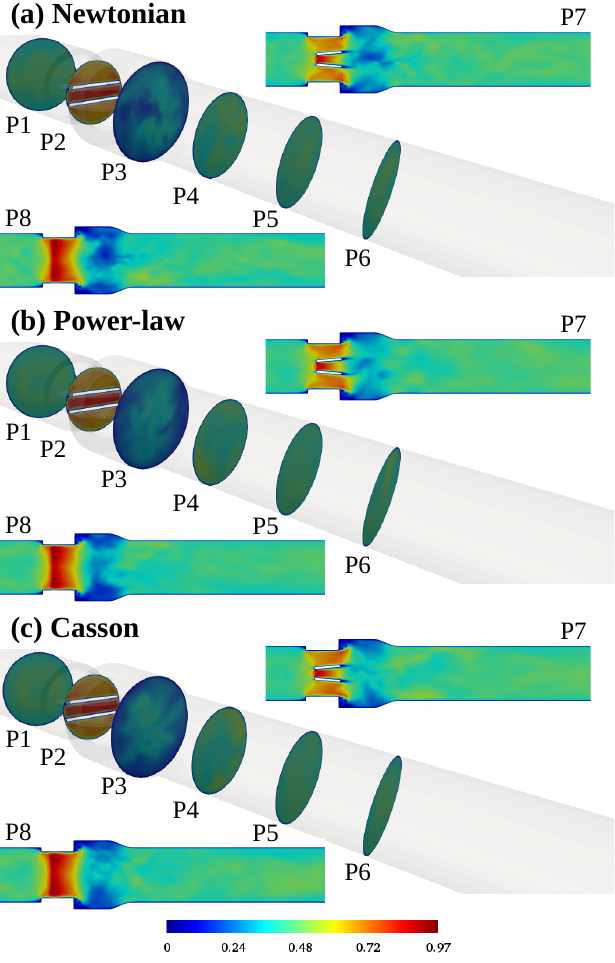}
    \caption{Velocity magnitude contours at eight different planes (P1-P8) during the mid-acceleration phase ($t = T1 \approx 1965$ ms) of the cardiac cycle predicted by Newtonian, power-law, and Casson fluid models of blood.} 
    \label{Umagt1}
\end{figure}

\begin{figure}
    \centering
    \includegraphics[trim=0cm 0cm 0cm 0cm,clip,width=8cm]{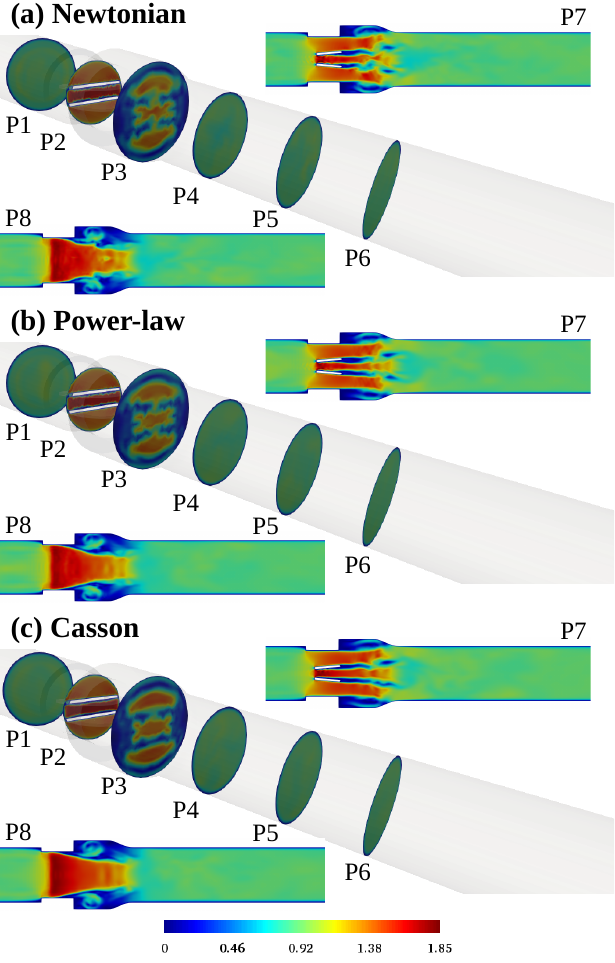}
    \caption{Velocity magnitude contours at eight different planes (P1-P8) during systolic peak ($t = T2 \approx 2074$ ms) of the cardiac cycle predicted by Newtonian, power-law, and Casson fluid models of blood.} 
    \label{Umagt2}
\end{figure}

\begin{figure}
   \centering
   \includegraphics[trim=0cm 0cm 0cm 0cm,clip,width=8cm]{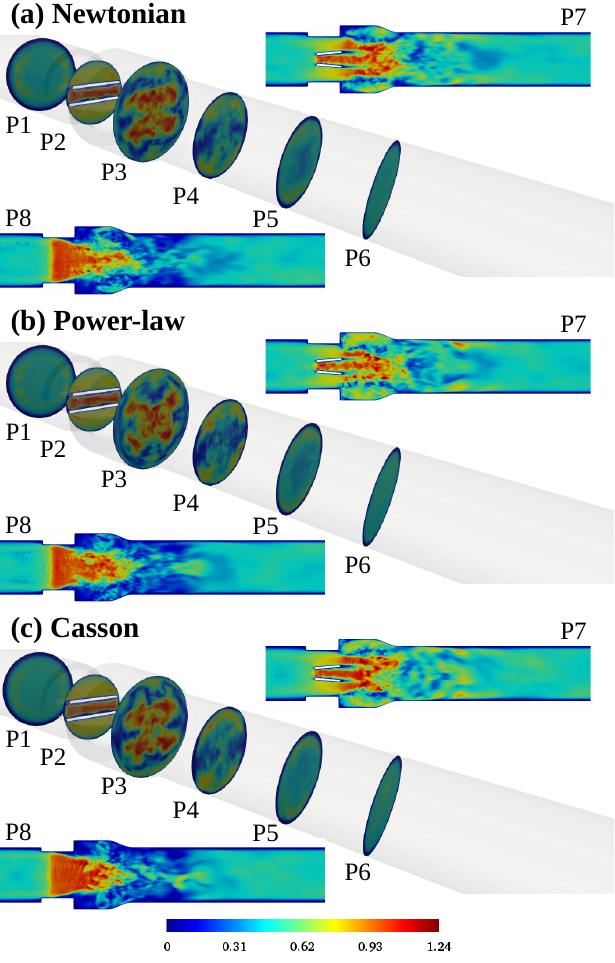}
   \caption{Velocity magnitude contours at eight different planes (P1-P8) during mid-deceleration phase ($t = T3 \approx 2170$ ms) of the cardiac cycle predicted by Newtonian, power-law, and Casson fluid models of blood.} 
    \label{Umagt3}
\end{figure}

\begin{figure*}
    \centering
    \includegraphics[trim=0cm 0cm 0cm 0cm,clip,width=16cm]{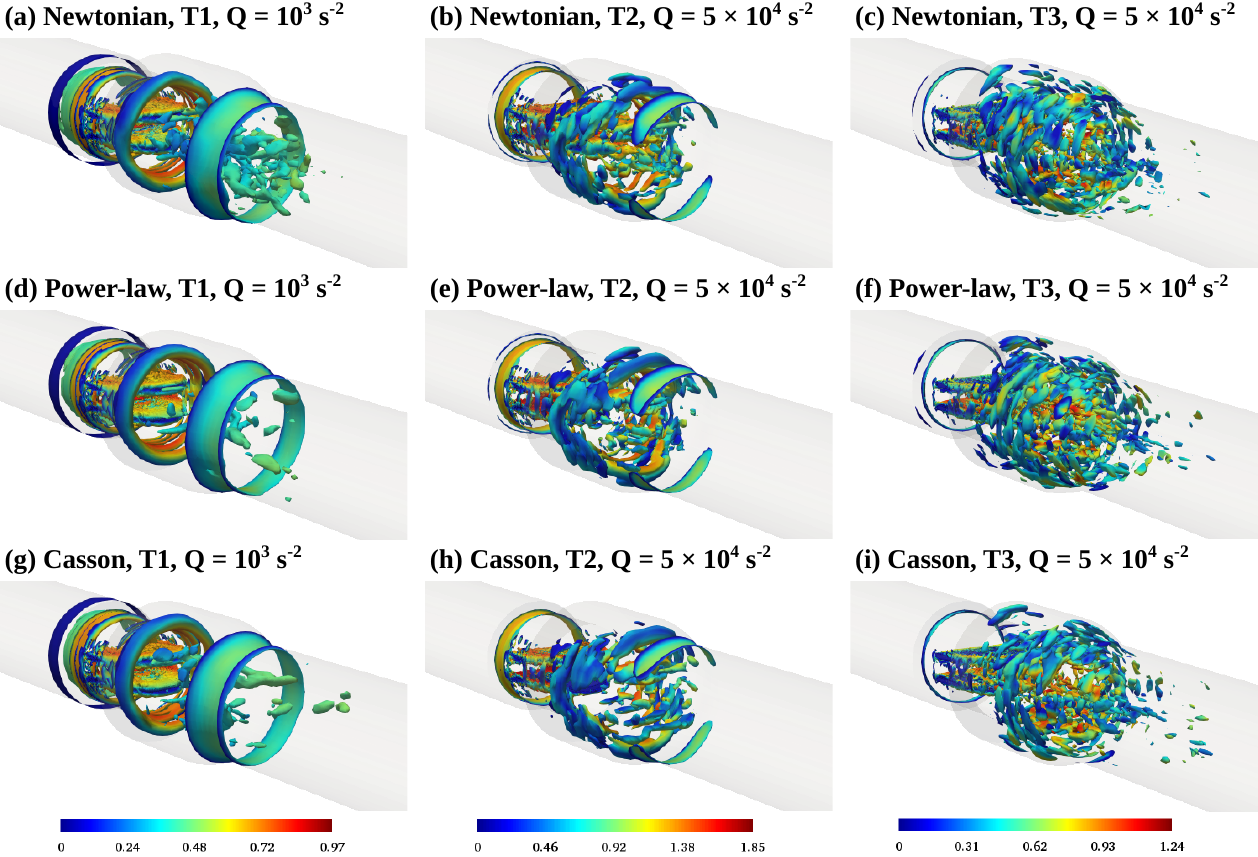}
    \caption{Isosurface for the Q-criteria during mid-acceleration ($t = T1 \approx 1965$ ms), peak systolic ($t = T2 \approx 2074$ ms) and mid-deceleration ($t = T3 \approx 2170$ ms) phase of the cardiac cycle predicted by Newtonian, power-law, and Casson fluid models of blood. The embedded color bar shows the corresponding magnitude of velocity.}
    \label{Q+Pulsatile}
\end{figure*}

\begin{figure*}
    \centering
    \includegraphics[trim=0cm 0cm 0cm 0cm,clip,width=16cm]{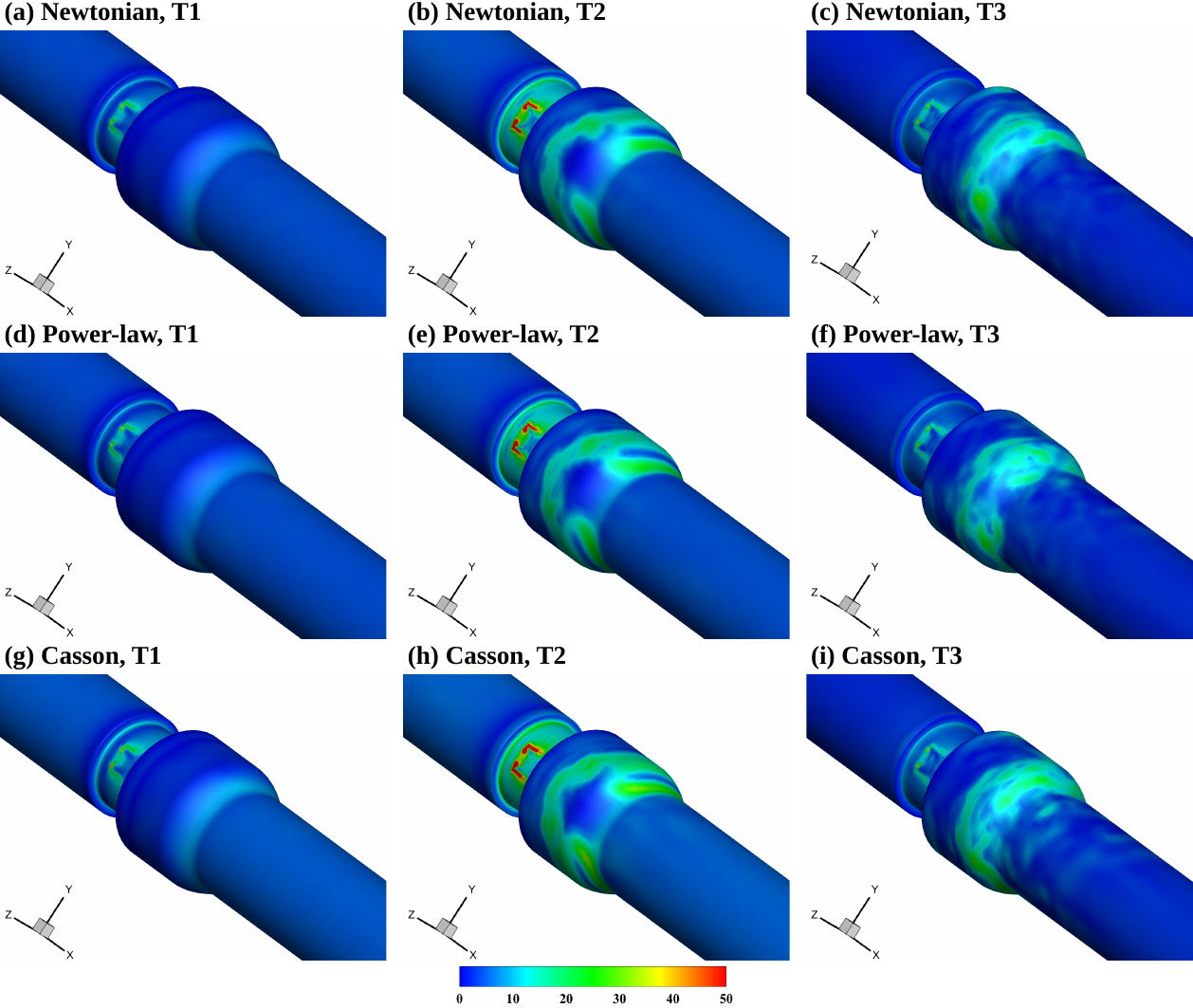}
    \caption{Variation of the wall shear stress (WSS) during mid-acceleration ($t = T1 \approx 1965$ ms), peak systolic ($t = T2 \approx 2074$ ms) and mid-deceleration ($t = T3 \approx 2170$ ms) phase of the cardiac cycle predicted by Newtonian, power-law, and Casson fluid models of blood. The values shown in the legend have units of `$Pa$'.}
    \label{WSSPulsatile}
\end{figure*}


\begin{figure*}
    \centering
    \includegraphics[trim=0cm 0cm 0cm 0cm,clip,width=16cm]{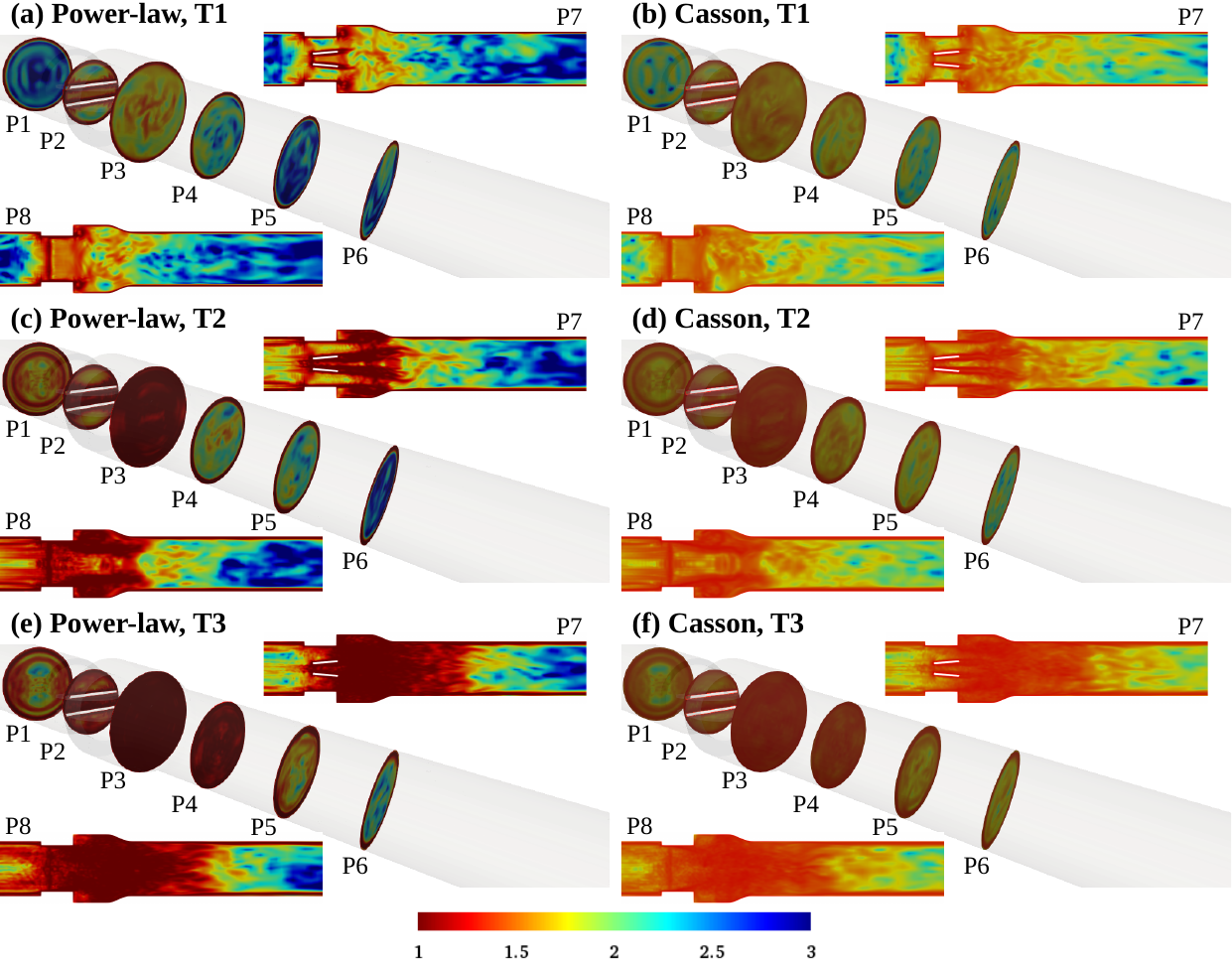}
    \caption{Local non-Newtonian importance factor ($I_L=\eta/\eta_{\infty}$) at eight different planes (P1-P8) during various time instances for the two non-Newtonian fluid models of blood, namely, (a) power-law, mid-acceleration phase ($t = T1 \approx 1965$ ms), (b) Casson, mid-acceleration phase ($t = T1 \approx 1965$ ms), (c) power-law, systolic peak ($t = T2 \approx 2074$ ms), (d) Casson, systolic peak ($t = T2 \approx 2074$ ms), (e) power-law, mid-deceleration phase ($t = T3 \approx 2170$ ms), and (f) Casson, mid-deceleration phase ($t = T3 \approx 2170$ ms).} 
    \label{ILPulsatile}
\end{figure*}

\begin{table}[hbt!]
  \caption{Values of various post-processed parameters for Newtonian, power-law, and Casson fluid models of blood obtained at three different time instances of the cardiac cycle, namely, mid-acceleration (T1), peak systolic (T2) and mid-deceleration (T3) phase.}
  \begin{center}
  \begin{ruledtabular}
  \begin{tabular}{ccccc}
    \textbf{Parameter} & {\textbf{Time instance}} & {\textbf{Newtonian}} & {\textbf{Power-law}} & {\textbf{Casson}} \\ [5pt]
    \multirow{3}{*}{\parbox{1.3cm}{~~| $\Delta P_{net}$| \\ ($mmHg$)}} & {T1} & 42.8346 & 42.8664 & 43.2321 \\
    & {T2} & 9.8898 & 9.9375 & 10.3350 \\
    & {T3} & 71.5476 & 71.3528 & 71.5238 \\[5pt]
    \multirow{3}{*}{\parbox{1.3cm}{~~$C_d$ \\ (Top leaflet)}} & {T1} & 28.4240 & 28.6189 & 30.4530 \\
    & {T2} & 66.7494 & 65.1499 & 69.0875 \\
    & {T3} & 7.2717 & 7.5626 & 7.7548 \\[5pt]
    \multirow{3}{*}{\parbox{1.3cm}{~~$C_d$ \\ (Bottom leaflet)}} & {T1} & 28.8118 & 28.5838 & 30.1858 \\
    & {T2} & 66.7503 & 66.2410 & 68.5901 \\
    & {T3} & 7.6539 & 7.5430 & 7.4631 \\[5pt]
    \multirow{3}{*}{\parbox{1.3cm}{$|WSS|^{max}_{walls}~$ \\ ($Pa$)}} & {T1} & 42.7490 & 42.5631 & 47.1142 \\
    & {T2} & 165.8206 & 167.3339 & 114.5817 \\
    & {T3} & 42.8277 & 39.7073 & 41.8817 \\[5pt]
    \multirow{3}{*}{\parbox{1.3cm}{$|WSS|^{avg}_{walls}~$ \\ ($Pa$)}} & {T1} & 4.3632 & 4.4072 & 5.0766 \\
    & {T2} & 5.4672 & 5.5287 & 6.3446 \\
    & {T3} & 3.0090 & 3.0660 & 3.4225 \\[5pt]
    \multirow{3}{*}{\parbox{1.3cm}{Blood damage ($\times 10^{6}$)}} & {T1} & $0.1384$ & $0.1383$ & $0.2102$ \\
    & {T2} & $0.5184$ & $0.5167$ & $0.7629$ \\
    & {T3} & $0.1738$ & $0.1783$ & $0.2617$ \\[2pt]
    \end{tabular}
    \end{ruledtabular}
  \label{table:Pulsatile}
  \end{center}
\end{table}

In the preceding section, steady inflow results have provided valuable insights into the flow physics associated with blood. However, it is the pulsatile nature of blood flow during the cardiac cycle that holds greater physiological importance. A pulsatile flow condition has been imposed at the inlet plane to accurately simulate this pulsatility, following a flow rate curve representative of one cardiac cycle (Fig.~\ref{FlowCurve}). This curve, spanning a time period of $860$ milliseconds, corresponds to a heart rate of $70$ beats per minute~\cite{Yun2014}. The simulation encompasses three cardiac cycles, with results analyzed at three different time instances: mid-acceleration (T1), peak systolic (T2), and mid-deceleration (T3) phases (Fig.~\ref{FlowCurve}). Additionally, the Casson fluid model has been incorporated alongside the power-law fluid model, which is widely used in studying blood hemodynamics. Newtonian results are also presented alongside the two non-Newtonian models for comparison under identical conditions. Figure~\ref{Umagt1} displays velocity magnitude contours at eight different planes (P1-P8) during the mid-acceleration phase (T1), representing the mid-acceleration phase of the cardiac cycle. Here, as blood reaches the leading edge of the valve leaflets, the flow separates into three jets, akin to steady inflow conditions. Moreover, the velocity magnitude of the central jet exceeds that of the lateral jets, regardless of the blood's rheological model. However, under this flow condition, all three jets are confined within the valve region only (region B in Fig.~\ref{Geometry}), in contrast to the steady inflow condition where they extend further downstream. The velocity field appears uniform along all planes normal to the x-axis except plane P3, situated in the sinus region (region C in Fig.~\ref{Geometry}), where flow separation occurs at the trailing edge of the leaflets, forming vortices, evident from the low velocity-magnitude values in this region. Conversely, the velocity field was highly non-uniform along all these planes under steady inflow conditions with nearly the same Reynolds number, Fig.~\ref{UmagSteady}. Additionally, the high-velocity magnitude zone in the valve region appears concave in shape (P8 plane), contrasting with the convex shape observed in steady inflow conditions. The velocity field is seen to be highly asymmetric in nature, signifying the existence of an unsteady flow field in the geometry at this flow condition.

As the velocity further increases during the peak systolic phase (T2), significant changes in flow behavior become apparent, Fig.~\ref{Umagt2}. Both the central and lateral jets extend in length and cover not only the valve region but also the sinus region. Beyond this, the jets disappear in the aortic region, and the velocity field becomes more uniform, as evident from the results presented in planes P4-P8. With the extension of the jets into the sinus region, a mushroom-shaped velocity field appears in plane P3, resembling observations in steady inflow conditions. However, the velocity magnitude is notably reduced within the sinus cavities, suggesting the potential formation of strong vortices in this region. Moreover, at this stage of the cardiac cycle, the lateral jets strengthen, becoming as robust as the central jet. Unlike at stage T1, where the central jet predominated, both the lateral jets and the central jet exhibit increased velocity magnitudes. This increase is consistent with the maximum flow rate and velocity expected during the peak systolic phase, resulting in the maximum expected jet length under these conditions.

In the mid-deceleration stage (T3), the velocity field exhibits greater chaotic behavior compared to the mid-acceleration (T1) and peak systolic (T2) stages, Fig.~\ref{Umagt3}. At this stage, the jets in the sinus region break apart and merge together, leading to a highly non-uniform velocity field, particularly noticeable in plane P3 within the sinus region. This non-uniformity extends up to plane P5 along the x-axis in the aortic region. The trend is further evident in the results presented on planes P7 and P8. This behavior during the cardiac cycle's mid-deceleration phase can be attributed to decreased flow rate and/or velocity. As the deacceleration phase progresses, fluid parcels with higher inertia slow down, increasing pressure in this region and creating an adverse pressure gradient. Consequently, the boundary layers separate more rapidly, leading to the breakage of jets and an escalation in chaotic behavior.

Figure~\ref{Q+Pulsatile} illustrates the isosurface of the Q-criterion at three distinct time instances during the cardiac cycle for three different rheological models of blood. The Q-criterion is mathematically defined as $Q = \frac{1}{2}\left( ||\bm{\Omega}||^{2} - ||\bm{S}||^{2} \right)$, where $\bm{\Omega} \left( = \frac{1}{2} \left( \nabla \bm{u} - \nabla \bm{u}^{T}\right)\right)$ and $\bm{S} \left( = \frac{1}{2} \left( \nabla \bm{u} + \nabla \bm{u}^{T}\right)\right)$ represent the vorticity and strain-rate tensors, respectively, and $||..||$ denotes their magnitudes. Positive values of the Q-criterion typically indicate the presence of vortex cores, representing concentrated vorticity with fluid particles rotating around an axis, indicative of coherent vortical structures within the flow. Conversely, negative values of the Q-criterion are associated with regions where strain dominates over rotation, leading to elongation or distortion of the flow instead of forming concentrated vortices. Thus, negative Q-values signify regions where the flow is more elongational than rotational. In Fig.~\ref{Q+Pulsatile}, the results depict two different values of Q = $1 \times 10^{3}$ and $5 \times 10^{4}$. At the mid-acceleration stage (T1) of the cardiac cycle, three prominent circumferential vortical structures emerge: around the trailing edge of the valve leaflets (valve region), the leading edge of the valve leaflets (sinus region), and at the beginning of the aortic section, irrespective of the blood rheological model. As the flow progresses to the peak systolic phase (T2), these large circumferential vortical structures fragment into smaller vortices, particularly in the sinus region. Consequently, the presence of small vortical structures increases significantly within the sinus and aortic regions, regardless of the rheological model, owing to the intensified flow at this stage. The tendency for further fragmentation into smaller vortical structures escalates as the flow enters the mid-deceleration phase (T3) of the cardiac cycle. Additionally, the extension of vortices into the aortic region expands at this stage. This is due to the increased chaos in the flow field resulting from the generation of an adverse pressure gradient, as observed in Fig.~\ref{Umagt3}, where the jets were observed to break due to the increased chaotic nature of the flow field. 

While the pattern and intensity of vortical structures appear qualitatively similar across all three rheological models of blood, notable differences also emerge. For instance, during the mid-acceleration stage (T1), small vortical structures manifest in the middle of the aortic section, with a more pronounced tendency observed for the Newtonian model (sub-Fig.~\ref{Q+Pulsatile}(a)) compared to the non-Newtonian power-law and Casson models. This difference may stem from the apparent viscosity of blood, which could be higher for the power-law and Casson models than for the Newtonian model during this stage of the cardiac cycle, owing to lower shear rates resulting from decreased flow rates and/or velocities. Consequently, the formation of small vortical structures for the non-Newtonian blood models may be suppressed. Furthermore, in the mid-deceleration stage of the cardiac cycle (T3), characterized by maximum chaos, the prevalence of small vortical structures is more pronounced for the power-law rheological model (sub-Fig.~\ref{Q+Pulsatile}(f)) compared to the Newtonian and Casson fluid models. This observation may be attributed to the shear-thinning behavior of blood captured by the power-law model, which promotes the formation of small vortical structures.

Figure~\ref{WSSPulsatile} illustrates the surface distribution of wall shear stress (WSS) at various stages of the cardiac cycle for different blood rheological models. In stage T1, WSS exhibits elevated values near the valve leaflets region, particularly in the valve hinge region situated within the valve housing section. As the cardiac cycle progresses to stage T2, the magnitude of WSS increases, with significantly higher values observed throughout the valve housing section than seen in stage T1. Additionally, patches of high WSS emerge in the sinus section during this stage. In the deceleration stage of the cardiac cycle (T3), the overall magnitude of WSS again decreases in the valve housing section, with regions of high WSS predominantly observed at the end of the sinus section. Although the distribution trend of WSS magnitude remains similar across all three blood rheological models, it appears higher and more widespread in the Casson model. For instance, in sub-Fig.~\ref{WSSPulsatile}(g) for stage T1, the results demonstrate this increased and broader distribution of WSS in the Casson model compared to the other two rheological models of blood.

The difference in spatial variations of flow characteristics, such as velocity magnitude or vortical structures among different rheological models of blood, is not prominently discernible. This observation is further elucidated by plotting the local non-Newtonian importance factor $I_{L}$ in Fig.~\ref{ILPulsatile} for the power-law and Casson rheological models at three distinct time instances during the cardiac cycle. As previously discussed, this factor quantifies the influence of non-Newtonian apparent viscosity on flow characteristics. In stage T1 of the cardiac cycle, $I_{L}$ values slightly exceed one near the valve leaflets and the wall region of the valve and sinus sections. This is due to the presence of high-shearing zones in these regions that reduce apparent viscosity (however, it remains lesser than the Newtonian one), resulting in $I_{L}$ values slightly above one. Conversely, in the aortic region, $I_{L}$ values are substantially higher than one throughout, except near the wall, for both power-law and Casson models. Progressing to stage T2, increased velocity (and hence higher shear rate) further diminishes apparent viscosity, with $I_{L}$ values precisely equal to one near the valve leaflets and valve and sinus regions for the power-law model. Additionally, the region of lower $I_{L}$ values extends deeper into the aortic region at this stage. In stage T3 of the cardiac cycle, this trend is accentuated. Notably, the Casson model exhibits a higher non-Newtonian importance factor than the power-law model across all stages of the cardiac cycle, indicating a greater impact on flow characteristics. 

Spatial variations in velocity magnitude, vortical structures, or wall shear stress offer valuable insights into the flow dynamics at different time instances during the cardiac cycle. However, it is the spatially averaged flow parameters that hold greater clinical significance. Hence, akin to steady inflow conditions, various post-processed spatially averaged parameters are presented in Table~\ref{table:Pulsatile} for different blood rheological models and cardiac cycle stages. From this table, it can be seen that the pressure drop is notably higher in stage T1 compared to T2, with a further increase observed in T3 surpassing T1 levels. Conversely, drag forces exerted on valve leaflets peak in T2 and reach a minimum in T3. These increased forces in the peak systolic stage stem from the maximum blood flow rate and velocity, consequently elevating shear stresses and frictional drag forces on the leaflet wall. As evident from the surface distribution of WSS, the corresponding maximum and surface-averaged wall shear stress values are highest in T2, as presented in Table~\ref{table:Pulsatile}. Notably, the Casson fluid model exhibits higher values for most parameters than the power-law and Newtonian models, which was also evident from the surface distribution of WSS in Fig.~\ref{WSSPulsatile}. For example, surface-averaged WSS values are 6.3446 and 5.4672 for the Casson and Newtonian models, respectively, implying a 16.4\% difference. Similarly, the Casson model shows a higher pressure drop, albeit with a 4.5\% difference compared to WSS. This discrepancy remains significant (>13\%) across other cardiac cycle stages, i.e., T1 and T3. This is due to the higher apparent viscosity inherent in the Casson model relative to the Newtonian counterpart. This trend aligns with observations from Fig.~\ref{ILPulsatile}, where the non-Newtonian importance factor shows the higher apparent viscosity associated with the Casson model, resulting in increased viscous forces and wall shear stresses. Moreover, the blood damage (BD) parameter exhibits its peak values when blood exhibits yield stress characteristics. Notably, the highest levels of damage occur during the T2 phase, regardless of the blood's non-Newtonian behaviors. A notable discrepancy in BD values arises between considering blood as a simple Newtonian fluid and as a Casson fluid, amounting to approximately 47\%, as shown in Table~\ref{table:Pulsatile}. This substantial difference signifies the importance of accounting for non-Newtonian blood behaviors, particularly yield stress, in the design of blood-contacting medical devices like mechanical heart valves, aiming to enhance their functionality and performance. 

\section{Conclusions}\label{Conclusions}
This study conducts extensive three-dimensional direct numerical simulations (DNS) to delve into the impact of non-Newtonian rheological behaviors of blood, including shear-thinning and yield stress, on the hemodynamics past a bileaflet mechanical heart valve. The simulations utilize the finite volume method (FVM) based on the open-source computational fluid dynamics (CFD) code OpenFOAM. Before diving into the main investigation, the accuracy and reliability of the present CFD solver are rigorously validated against existing experimental and numerical data from the literature, ensuring the robustness of the present computational framework. Simulations are conducted under both steady inflow and physiologically realistic pulsatile flow conditions. In steady inflow conditions, three central jets form—one between the two valve leaflets and two lateral jets between the leaflets and the valve housing wall, which decrease in size with the Reynolds number. A mushroom-like flow structure emerges downstream of the valve leaflets, gradually diminishing with increasing Reynolds number. Furthermore, with the increasing values of the Reynolds number, a gradual transition in the flow field from a steady and laminar to an unsteady and turbulent is observed. Under pulsatile flow conditions, jet sizes increase from the mid-acceleration stage to the peak stage of the cardiac cycle, ultimately dissipating in the deceleration stage due to adverse pressure gradients. At this stage, small vortical structures appear downstream of the valve leaflets, although jet sizes remain smaller than those observed in steady inflow conditions.

While the spatial variations in the flow field due to blood non-Newtonian behaviors may not be immediately apparent, their influence on surface and time-averaged clinical quantities like wall shear stress (WSS), pressure recovery, and blood damage is substantial. For instance, WSS values are significantly higher in the shear-thinning power-law model compared to the Newtonian model, with a difference of around 67\% at $Re = 750$, gradually decreasing to less than 1\% at $Re = 5000$. Similarly, pressure recovery downstream of the valve leaflets is lower in the presence of the shear-thinning behavior of blood. Additionally, WSS values are increased under pulsatile flow conditions, especially when considering shear-thinning and yield stress behaviors of blood, with yield stress behavior exhibiting over 13\% higher values than the Newtonian model. Furthermore, drag forces acting on the valve leaflets are increased when non-Newtonian blood behaviors are considered, indicating their significant influence on hemodynamics and associated consequences such as blood cell damage. It is also evident in the blood damage (BD) parameter values, showing a variation of approximately 47\% between blood exhibiting non-Newtonian yield stress characteristics and blood characterized by a constant Newtonian viscosity.

However, it should be emphasized here that this study assumes fixed leaflets at the maximum opening, omitting fluid-structure interaction (FSI) that occurs in real scenarios during mechanical heart valve operation. Additionally, the valve housing wall is treated as a rigid solid, whereas in reality, it is flexible and extensible. Addressing these aspects would offer a more realistic depiction of blood non-Newtonian behaviors' influence on the corresponding hemodynamics. However, addressing all these aspects would require solving additional physics, which, in turn, would necessitate more computational hours, which will be explored in our subsequent studies.

\section{Acknowledgements}
We acknowledge the National Supercomputing Mission
(NSM) for providing computing resources of ‘PARAM Smriti’ at NABI, Mohali (accessed by CS) and ‘PARAM Himalaya’ at IIT Mandi (accessed by AC), which are implemented by C-DAC and supported by the Ministry of Electronics and Information Technology (MeitY) and Department of Science and Technology (DST), Government of India. AC would also like to thank the Ministry of Education, Government of India, for the financial support provided by the PMRF (Cycle-9). CS acknowledges the Indian Council of Medical Research (ICMR) and Science and Engineering Research Board (SERB), Government of India, for the financial support provided through the grant numbers 17X(3)/Adhoc/17/2022-ITR and ECR/2018/000202, respectively.


\section{References}
\bibliography{aipsamp}

\nocite{*}
\end{document}